\documentclass[twocolumn]{aastex631}
\usepackage{graphicx}
\usepackage{url}
\usepackage{epstopdf}
\usepackage{makecell}
\usepackage{xcolor}
\usepackage{textcomp}
\usepackage{gensymb}
\usepackage{multirow}
\usepackage{ragged2e}
\usepackage{hyperref}
\usepackage{amsmath}
\usepackage{booktabs}
\usepackage{comment}
\usepackage{afterpage}
\usepackage{mathtools}
\usepackage{tabularx}
\usepackage{bold-extra}
\usepackage{xspace}
\usepackage{relsize}
\usepackage[utf8]{inputenc}
\usepackage{newunicodechar}
\usepackage[encapsulated]{CJK}
\usepackage{ucs}
\usepackage{todonotes}
\usepackage{threeparttable}
\usepackage{nicefrac}
\usepackage[T1]{fontenc}
\usepackage{rotating}

\def\sgra{Sgr~A$^*$\xspace}
\def\m87{M87$^*$\xspace}

\begin{document}

\title{Accessing a New Population of Supermassive Black Holes with Extensions to the Event Horizon Telescope}

\author[0009-0007-5412-1894]{Xinyue Alice Zhang}
\affiliation{Center for Astrophysics $\vert$ Harvard \& Smithsonian, 60 Garden Street, Cambridge, MA 02138, USA}
\affiliation{Department of Physics, Harvard University, 17 Oxford Street Cambridge, MA 02138, USA}
\affiliation{Black Hole Initiative, 20 Garden Street, Cambridge, MA 02138, USA}

\author[0000-0001-5287-0452]{Angelo Ricarte}
\affiliation{Center for Astrophysics $\vert$ Harvard \& Smithsonian, 60 Garden Street, Cambridge, MA 02138, USA}
\affiliation{Black Hole Initiative, 20 Garden Street, Cambridge, MA 02138, USA}

\author[0000-0002-5278-9221]{Dominic W. Pesce}
\affiliation{Center for Astrophysics $\vert$ Harvard \& Smithsonian, 60 Garden Street, Cambridge, MA 02138, USA}
\affiliation{Black Hole Initiative, 20 Garden Street, Cambridge, MA 02138, USA}

\author[0000-0002-4120-3029]{Michael D. Johnson}
\affiliation{Center for Astrophysics $\vert$ Harvard \& Smithsonian, 60 Garden Street, Cambridge, MA 02138, USA}
\affiliation{Black Hole Initiative, 20 Garden Street, Cambridge, MA 02138, USA}

\author[0000-0001-6920-662X]{Neil Nagar}
\affiliation{Astronomy Department, Universidad de Concepci\'on, Casilla 160-C, Concepci\'on, Chile}

\author[0000-0002-1919-2730]{Ramesh Narayan}
\affiliation{Center for Astrophysics $\vert$ Harvard \& Smithsonian, 60 Garden Street, Cambridge, MA 02138, USA}
\affiliation{Black Hole Initiative, 20 Garden Street, Cambridge, MA 02138, USA}

\author[0000-0002-9248-086X]{Venkatessh Ramakrishnan}
\affiliation{Finnish Centre for Astronomy with ESO, FI-20014 University of Turku, Finland}
\affiliation{Aalto University Mets\"ahovi Radio Observatory, Mets\"ahovintie 114, FI-02540 Kylm\"al\"a, Finland}

\author[0000-0002-9031-0904]{Sheperd Doeleman}
\affiliation{Center for Astrophysics $\vert$ Harvard \& Smithsonian, 60 Garden Street, Cambridge, MA 02138, USA}
\affiliation{Black Hole Initiative, 20 Garden Street, Cambridge, MA 02138, USA}

\author[0000-0002-7179-3816]{Daniel C. M. Palumbo}
\affiliation{Center for Astrophysics $\vert$ Harvard \& Smithsonian, 60 Garden Street, Cambridge, MA 02138, USA}
\affiliation{Black Hole Initiative, 20 Garden Street, Cambridge, MA 02138, USA}

\correspondingauthor{Xinyue Alice Zhang}

\begin{abstract}

The Event Horizon Telescope has produced resolved images of the supermassive black holes \sgra and \m87, which present the largest shadows on the sky.  In the next decade, technological improvements and extensions to the array will enable access to a greater number of sources, unlocking studies of a larger population of supermassive black holes through direct imaging. In this paper, we identify 12 of the most promising sources beyond \sgra and \m87 based on their angular size and millimeter flux density. For each of these sources, we make theoretical predictions for their observable properties by ray tracing general relativistic magnetohydrodynamic models appropriately scaled to each target's mass, distance, and flux density.  We predict that these sources would have somewhat higher Eddington ratios than \m87, which may result in larger optical and Faraday depths than previous EHT targets.  Despite this, we find that visibility amplitude size constraints can plausibly recover masses within a factor of 2, although the unknown jet contribution remains a significant uncertainty.  We find that the linearly polarized structure evolves substantially with Eddington ratio, with greater evolution at larger inclinations, complicating potential spin inferences for inclined sources.  We discuss the importance of 345 GHz observations, milli-Jansky baseline sensitivity, and independent inclination constraints for future observations with upgrades to the Event Horizon Telescope (EHT) through ground updates with the next-generation EHT (ngEHT) program and extensions to space through the Black Hole Explorer (BHEX). 
\end{abstract}


\section{Introduction} \label{sec:intro}

The Event Horizon Telescope (EHT) collaboration has produced resolved images of two supermassive black holes (SMBHs), \m87 \citep{EHTC+2019a,EHTC+2019b,EHTC+2019c,EHTC+2019d,EHTC+2019e,EHTC+2019f,EHTC+2021a,EHTC+2021b,EHTC+2023,EHTC+2024a,EHTC+2025} and \sgra \citep{EHTC+2022a,EHTC+2022b,EHTC+2022c,EHTC+2022d,EHTC+2022e,EHTC+2022f,EHTC+2024b,EHTC+2024c}.  This was accomplished using Very Long Baseline Interferometry (VLBI) at a frequency of 230 GHz, which, limited by the size of the Earth, achieved a resolution of approximately 20 $\mu$as.  These images confirmed the existence of a black hole's apparent shadow and have enabled new tests of gravity and accretion physics.

A number of extensions to the array are planned to produce higher quality images and movies of these two SMBHs, as well as enable demographic studies for a larger population of SMBHs.  The next-generation Event Horizon Telescope (ngEHT) project aims to add several new dishes around the world while upgrading many of the existing EHT sites \citep{Doeleman+2023,Johnson+2023}.  These upgrades would enable simultaneous multi-frequency (86-230-345 GHz) observations, substantially improving the array sensitivity, as well as expanded observing campaigns that will be capable of producing movies that image the accretion disk and jet simultaneously with high dynamic range.  Through observations at 345GHz and increased baseline coverage, the ngEHT will greatly improve the angular resolution possible with ground VLBI arrays.  Meanwhile, the Black Hole Explorer (BHEX) aims to complement the ground array with an orbiting satellite \citep{Johnson+2024}.  The higher angular resolution provided by space baselines will enable photon ring detections of \m87 and potentially also \sgra.  Both ngEHT and BHEX will allow access to black hole shadows with smaller angular sizes on the sky.

These array improvements will resolve horizon scale emission for a greater population of sources, enabling demographic studies for the first time. Theoretical studies of the EHT images of both \sgra and \m87 favor models with dynamically important magnetic fields \citep[so-called ``magnetically arrested disks'' or MADs;][]{Bisnovatyi-Kogan&Ruzmaikin1974,Igumenshchev+2003,Narayan+2003}.  MAD models power efficient jets \citep{Tchekhovskoy+2011} via the \citet[][]{Blandford&Znajek1977} (BZ) mechanism, which has important implications for SMBH-galaxy co-evolution.  With a larger sample of objects, we can begin to test the universality of this accretion state and investigate trends as a function of SMBH mass, inclination, Eddington ratio, and host galaxy properties.  The morphology of the linear polarization has also been proposed as a probe of spin, since synchrotron emission inherits the geometry of the magnetic field, which in turn is affected by frame dragging on event horizon scales \citep{Palumbo+2020,Emami+2023,Ricarte+2022,Chael+2023}, although there remain theoretical uncertainties with respect to the electron thermodynamics and Faraday effects. Potential spin constraints with extensions to the EHT could be used to test the Blandford-Znajek mechanism as well as models of SMBH assembly over cosmic time \citep{Ricarte+2023a}.

By modeling the SMBH mass function and Eddington ratio distribution, \citet{Pesce+2021} estimated that the ngEHT could access event horizon scale structure of tens more sources.  To identify real candidates, the ongoing Event Horizon and Environs (ETHER) survey has been pursuing millimeter observations of promising sources and assembling a database of multi-frequency observations \citep{Ramakrishnan+2023,Hernandez+2024, Nair+2024}.  \citet{Pesce+2022} later simulated futuristic array performance and confirmed that the ngEHT could plausibly measure masses (from inferred ring diameters) and spins (from inferred polarized morphology) for tens of sources using geometric modeling.  However, it remained to be tested if such inferences would be valid for physical models with realistic Eddington ratios and inclinations.

In this work, using the ETHER database as a starting point, we perform the first in-depth case study of 12 of the (currently identified) most promising targets for the EHT as well as its extensions through the ngEHT and BHEX, using general relativistic magnetohydrodynamics (GRMHD) simulations followed by polarized general relativistic ray-tracing (GRRT).  A similar list was assembled prior to EHT observations in \citet{Johannsen+2012}, and since then, available data on SMBH mass measurements, their spectral energy distributions, array specifications, and simulation tools have advanced substantially.  In \autoref{sec:targets}, we study the ETHER catalogue and describe these 12 most promising candidate sources.  In \autoref{sec:GRMHD}, we produce a new library of simulated images of these sources from general relativistic fluid and radiative transfer simulations, which we use to infer Eddington ratios and make predictions for demographic science.  We discuss and summarize our findings in \autoref{sec:sum}.

\section{Summary of Future Targets} 
\label{sec:targets}

We begin our investigation using the ETHER database, which compiles a large quantity of masses, multi-frequency flux measurements, and host galaxy information for 3.8 million sources \citep{Ramakrishnan+2023,Hernandez+2024, Nair+2024}.  
\citet{Pesce+2022} determined a minimum 230 GHz flux density as a function of angular size required to make mass measurements of optically thin black holes by the ngEHT\footnote{Note that for simplicity, we refer to the EHT upgraded as described in \autoref{sec:intro} as the ngEHT throughout this paper.}.  We select sources by hand from ETHER that satisfy this criterion based on currently available data. We note that 230 GHz flux measurements or predictions of ETHER sources with large angular sizes are continually being added to the database, so the ngEHT-observable sample is expected to grow significantly over the upcoming years.  The relevant angular size is that of the photon ring, which for a non-spinning black hole, has an angular diameter given by

\begin{equation}
    \theta = 2\sqrt{27}\frac{GM_\bullet}{c^2D}
    \label{eqn:photon_ring}
\end{equation}

\noindent where $G$ is the gravitational constant, $M_\bullet$ is the SMBH mass, $c$ is the speed of light, and $D$ is the distance to the object.  If the SMBH is spinning, the photon ring changes size by only $\sim$10\% \citep[e.g.,][]{Johannsen&Psaltis2010,Johnson+2020,Chael+2021}.

Most of the mass estimates in this work come from direct gas or stellar dynamical modeling.  Note that we only adopt a single value of SMBH mass, distance, and mm flux density for each source, but systematic error dominates the error in these quantities.  For most of our sources, there are multiple mass estimates arising from different datasets and different methodologies.  For the sources in our sample, the average difference between the maximum and minimum mass estimate is 0.8 dex, and the average standard deviation among different mass estimates for the same object is 0.3 dex.

For each object, we adopt a 230 GHz flux density based on available data.  These values are used as inputs for our ray-tracing calculations, which are normalized to reproduce the appropriate flux density (see \autoref{sec:GRMHD}).  Only for \sgra and \m87 can we utilize estimates of the compact flux density from the EHT.  For the rest of our sources, the unknown compact flux fraction is a key systematic uncertainty in our work.  For \m87, the compact flux density is about half the flux density measured by ALMA alone \citep{EHTC+2019a}.  For \sgra, EHT constrains at least 80\% of the observed flux to the compact region \citep{EHTC+2022c}.  Lacking available 230 GHz VLBI constraints for the rest of our sources, we assume that 100\% of the flux density observed at 230 GHz, by single dish or phased-array observations as described for each source, can be attributed to the compact (non-jet) region modeled by our general relativistic magnetohydrodynamics GRMHD simulations.  Due to the incompleteness of 230 GHz flux density measurements, we adopted the 345 GHz flux density for NGC 4552, the 43 GHz flux density for NGC 3998, and the 8.4 GHz flux density for NGC 2663, effectively assuming an spectral index of 0.

As a spot check, we also refitted $\mathcal{M}$ and inferred Eddington ratios assuming that only either 50\% or 10\% of the total flux density can be attributed to the compact component which we simulate.  In these cases, the Eddington ratios decreased to 63\% and 23\% of their original values on average.  This sub-linear dependence of the Eddington ratio of compact flux density is expected, following the scalings described in \autoref{sec:eddington_ratios}.\footnote{Near the synchrotron emission peak, the emissivity coefficient $j_\nu \propto n_eB^2$.  Since $n_e\propto f_\mathrm{Edd}$ and $B \propto \sqrt{f_\mathrm{Edd}}$, we expect that $f_\mathrm{Edd} \propto \sqrt{F_{230}}$.  Therefore, if our flux densities were over-estimated by a factor of 10, the Eddington ratio would only be over-estimated by a factor of 3.} In the future, better estimates of the compact 230 GHz flux will be obtained either directly, by ongoing EHT observation programs, or by SED modeling \citep[see e.g.,][]{Bandyopadhyay+2019}.

Our sample includes a variety of black hole masses, stellar masses, morphologies, and environments.  In \autoref{appendix:properties}, we summarize some interesting characteristics of these objects, listed in descending projected shadow size.  The types of objects accessible to extensions of the EHT are similar to the types of objects most likely to source nHz gravitational waves (massive, nearby SMBHs) \citep{Mingarelli+2017}, which may potentially enable multi-messenger science.  Almost all of these objects have jets from which inclination constraints can potentially be obtained.  We reiterate that this is not an exhaustive list of objects observable by ngEHT and/or BHEX, and that it is simply a tractable sample based on currently available data.

\begin{sidewaystable}[t]
\begin{center}
\begin{threeparttable}
\vspace{9cm}
\hbox to \textwidth{
  \hspace*{-1.6in}
  \begin{tabular}{|p{1.5 cm} p{1 cm} p{1 cm} p{1.5 cm} p{1.2 cm} p{1.3 cm} p{1.5 cm} p{1.1 cm} p{1.5 cm} p{1.2 cm} p{1.4 cm} p{1.3 cm} p{1.2 cm} p{1.75 cm} | }
    \hline
Galaxy & RA [h] & Dec [$^\circ$] & Distance [Mpc] & log($M_\bullet$ [$M_\odot$]) & Mass Type & Galaxy Morphology & Radio Loud /Quiet & mm Flux Density [Jy] & Shadow size [$\mu$as] & log(Stellar Mass [$M_\odot$]) & AGN Type & Jet orientation [$^\circ$] & Model $f_\mathrm{Edd}$ [$10^{-6}$] (This Work) \\
    \hline
SgrA* & 17.1 & -29.0 & $0.008 \pm 0.00004$ & $6.6^{+0.0014}_{-0.0014}$ & Stellar & Spiral & RL & $2.4 \pm 0.28$ & 54 & 10.8 & LLAGN &  & 0.035 \\
M87 & 12.5 & 12.4 & $17 \pm 0.62$ & $9.8^{+0.026}_{-0.027}$ & Stellar & Elliptical & RL & $0.50 \pm 0.05$ & 38 & 11.5 & LLAGN & $17.2 \pm 3.3$ & 0.52 \\
IC 1459 & 23.0 & -36.5 & $29 \pm 3.7$ & $9.4^{+0.077}_{-0.035}$ & Stellar & Elliptical & RL & $0.22 \pm 0.021$ & 8.9 & 10.8 & LLAGN &  & 1.48 \\
NGC 4594 & 12.7 & -11.6 & $9.9 \pm 0.82$ & $8.8^{+0.025}_{-0.027}$ & Stellar & Spiral & RQ & $0.20 \pm 0.01$ & 7.0 & 11.3 & LIN, LLAGN & $66^{+4}_{-6}$ & 1.12 \\
NGC 3998 & 12.0 & 55.5 & $14 \pm 1.3$ & $8.9^{+0.035}_{-0.035}$ & Stellar & Lenticular & RQ & $0.13 \pm 0.024$ & 6.2 & 10.2 & LIN, LLAGN, Sy1 & $10$–$21$ & 1.08 \\
NGC 4261 & 12.3 & 5.8 & $31 \pm 1.6$ & $9.2^{+0.11}_{-0.099}$ & CO & Elliptical & RQ & $0.20 \pm 0.05$ & 5.6 & 10.7 & LIN, LLAGN & $63 \pm 3$ & 2.57 \\
NGC 2663 & 8.8 & -33.8 & $28 \pm 1.4$ & $9.2^{+0.58}_{-0.61}$ & M–$\sigma$ & Elliptical & RL & $0.084 \pm 0.0084$ & 5.4 & 11.7 &  &  & 1.18 \\
NGC 3894 & 11.8 & 59.49 & $50 \pm 2.5$ & $9.4^{+0.48}_{-0.48}$ & M–$\sigma$ & Elliptical & RL & $0.058 \pm 0.015$ & 5.4 & 10.7 & LIN, LLAGN &  & 1.19 \\
M84 & 12.4 & 12.9 & $19 \pm 0.60$ & $9.0^{+0.042}_{-0.043}$ & Gas & Elliptical / Lenticular & RL & $0.13 \pm 0.015$ & 5.2 & 11.4 & LLAGN & $58^{+17}_{-18}$ & 1.52 \\
NGC 4552 & 12.6 & 12.6 & $15 \pm 0.99$ & $8.7^{+0.051}_{-0.051}$ & Stellar & Elliptical & RL & $0.027 \pm 0.009$ & 3.4 & 10.3 & LINER, LLAGN &  & 0.63 \\
3C 317 & 15.3 & 7.0 & $140 \pm 7.2$ & $9.7^{+0.022}_{-0.032}$ & M–$L_\mathrm{bulge}$ & Elliptical & RL & $0.034 \pm 0.002$ & 3.3 & 11.2 & LLAGN &  & 2.18 \\
NGC 315 & 1.0 & 14.5 & $70 \pm 3.5$ & $9.3^{+0.064}_{-0.033}$ & CO & Elliptical & RL & $0.18 \pm 0.009$ & 3.1 & 11.2 & LLAGN &  & 8.89 \\
NGC 1218 & 3.1 & 4.1 & $120 \pm 5.9$ & $9.5^{+0.48}_{-0.48}$ & M–$\sigma$ & Lenticular & RL & $0.11 \pm 0.01$ & 3.0 & 11.1 & BL Lac, Sy1 &  & 7.36 \\
NGC 5077 & 13.3 & -12.7 & $39 \pm 8.4$ & $8.9^{+0.18}_{-0.32}$ & Gas & Lenticular & RL & $0.068 \pm 0.007$ & 2.3 & 10.8 & LIN, LLAGN &  & 3.91 \\
\hline
\end{tabular}
}
\caption{Properties of potential horizon-scale targets for an Enhanced EHT adopted for or inferred from this study, listed in order of descending projected shadow size. Distance and mm flux density, with the exception of \sgra and \m87, are taken from the ETHER Database (Nagar et al., in prep.). Radio-loudness threshold is taken from \citet{Wang+2024}, where $\log(L_R/L_X)>-2.73$, with $L_R$ the 5 GHz luminosity from the RFC, NED, and \citet{Helmboldt+2007}, and $L_X$ the 2–10 keV luminosity from the Chandra Source Catalog and NED. Shadow sizes are computed analytically from $M_\bullet$ and distance. Stellar masses (except \sgra) are from WISE photometry \citep{Hernandez+2024}.}
\label{table:sources}
\end{threeparttable}
\end{center}
\end{sidewaystable}

Several key properties of our sources are summarized in \autoref{table:sources}. While most reside in elliptical galaxies, several of them exhibit lenticular morphologies and one of them is a spiral. These galaxies are also located in a range of intergalactic environments: 3C 317 is a brightest cluster galaxy (BCG), NGC 315, NGC 4261, and M84 are located within clusters, while IC 1459 is in a loose galaxy group.  We provide radio-loud (RL) versus radio-quiet (RQ) classifications using the definition of \citet{Wang+2024}, who propose a division between the populations at $\log(L_R/L_X)>-2.73$, where $L_R$ is the flux measured in the 5 GHz band and $L_X$ is the hard X-ray flux.  Our sample contains both types of object, and it will be very interesting to test accretion models as a function of radio-loudness on event-horizon scales.  The most promising candidates are those with large black hole shadow size and millimeter flux density. Excluding \sgra and \m87 which were observed by EHT, they are IC 1459, NGC 4594, and NGC 4261.  

Note that while our sources contain LINERs and several Seyferts, whose emission lines are characterized by cooler species, the existence of low temperature emission lines do not rule out the existence of an advection dominated accretion flow (ADAF) in the innermost accretion regions. In fact, it is expected that an inner ADAF should exist in all but the highest Eddington ratio AGN \citep[e.g.,][]{Esin+1997,Narayan&McClintock2008,Yuan&Narayan2014}.  These predictions remain to be tested via millimeter VLBI observations on event horizon scales.

The recently reported ``mm fundamental plane of accretion'' offers a sanity check for our flux density values \citep{Ruffa+2024}. This new fundamental plane relation differs from that of \citet{Falcke+2000,Merloni+2003} in using 5 GHz instead of 230 GHz. This is an empirical correlation between the nuclear 1 mm luminosity ($L_{\nu,\text{mm}}$), the intrinsic 2-10 keV X-ray luminosity ($L_{\text{X, 2-10}}$), and the SMBH mass ($M_{\bullet}$), given by:

\begin{align}
\begin{split}
    \log_{10}\left(\frac{M_\bullet}{M_{\odot}}\right) = & \;(-0.23\pm 0.05)\left[ \log_{10}\left(\frac{L_{\text{X, 2-10}}}{\text{erg s}^{-1}}\right)-40 \right] \\
    & +(0.95\pm 0.07)\left[ \log_{10}\left(\frac{L_{\nu\text{, mm}}}{\text{erg s}^{-1}}\right)-39 \right] \\
    & +(8.35\pm 0.08)
\end{split}
\end{align} 
 
$M_\bullet$ also significantly correlates solely with $L_{\nu,\text{mm}}$, although the correlation is tighter when $L_{\text{X, 2-10}}$ is included:

\begin{align}
\begin{split}
    \log_{10}\left(\frac{M_\bullet}{M_{\odot}}\right) = & \;(0.79\pm 0.08)\left[ \log_{10}\left(\frac{L_{\nu,\text{mm}}}{\text{erg s}^{-1}}\right)-39 \right] \\
    & +(8.2\pm 0.1)
\end{split}
\end{align}

\begin{figure*}[htb]
\centering
\includegraphics[width=\textwidth]{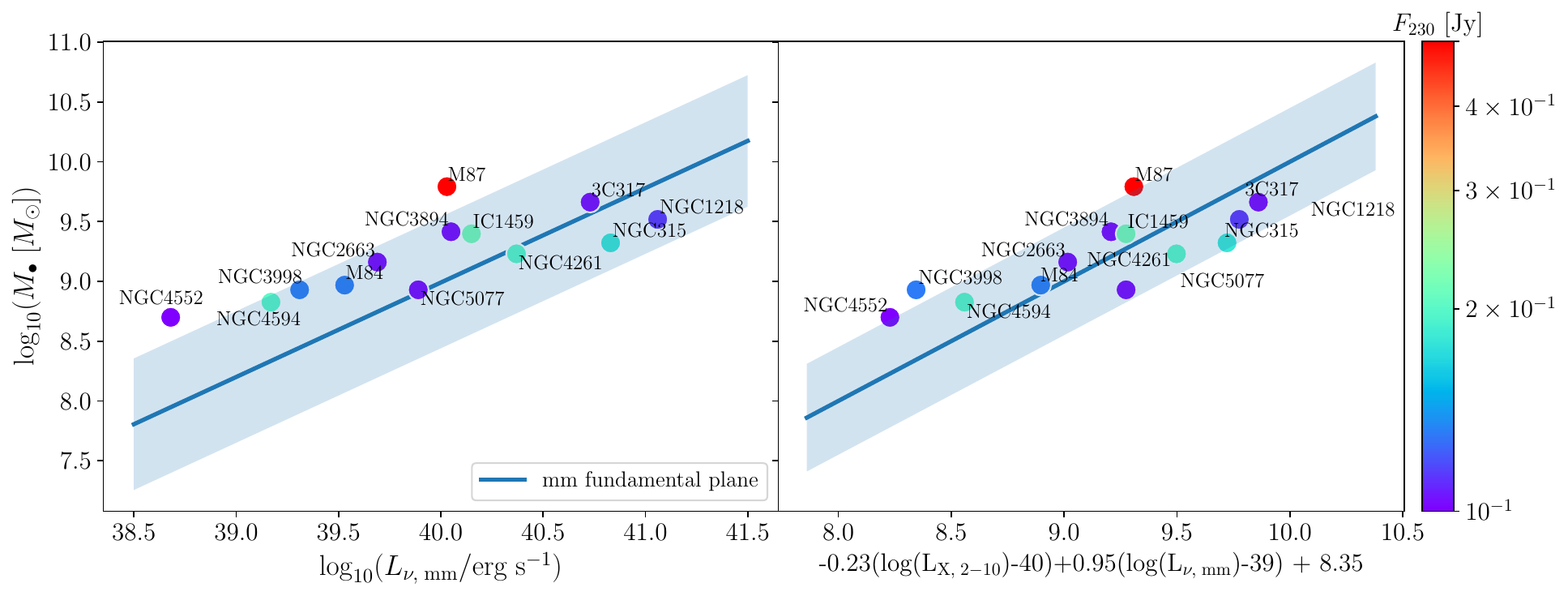}
\caption{Sources in our sample compared to the ``mm fundamental plane'' reported by \citet{Ruffa+2024} (blue band), relating SMBH mass ($M_\bullet$), mm luminosity ($L_{\nu, \mathrm{mm}}$), and hard X-ray luminosity ($L_{\mathrm{X},2-10}$). SMBH mass and mm luminosity are taken from the ETHER Database (Nagar et al., in prep.), and hard X-ray luminosity is taken from the Chandra Source Catalog \citep{Evans2010} and NED. In this and subsequent figures, symbol color encodes the 230 GHz flux density.  Our sample is consistent with the fundamental plane, confirming that our sources have typical masses and mm flux densities.} 
\label{fig:mmfundplane}
\end{figure*}

\autoref{fig:mmfundplane} shows that our sources\footnote{\sgra has much lower mass, higher flux density, and lower Eddington ratio compared to the rest of our objects. Since it is an outlier and its status as an EHT target is well-established, it is omitted from most of our figures.} are consistent with the mm fundamental plane.  The left panel shows the correlation between $M_\text{BH}$ and $L_{\nu,\text{mm}}$ and the right panel shows the correlation between $M_\text{BH}$, $L_{\nu,\text{mm}}$, and $L_{\text{X, 2-10}}$. All values of $L_{\mathrm{X},2-10}$ are taken from \cite{Gonzalez-Martin2009}, with the exception of NGC 5077, taken from \cite{gultekin+2012} and 3C 317, taken from \cite{Mezcua+2018}. For NGC 2663, NGC 3894, and NGC 1218, $L_{\mathrm{X},2-10}$ is currently unavailable in literature. Our sources are broadly consistent with this relation, but may be biased mildly towards larger masses at the low-luminosity end, since we select for larger shadow sizes.

\section{Generating Model Images with GRMHD}\label{sec:GRMHD}

There had long been many predictions of the properties of spatially-resolved horizon-scale images \citep[e.g.][]{Falcke+2000, Broderick+2009}. In this work, using mass, distance, and 230 GHz flux density, we generate astrophysical model images from SMBH accretion disk simulations \citep[following, the Patoka pipeline][]{Wong+2022}.  We start with the MAD GRMHD simulations of \citet{Narayan+2022}, performed using the code {\sc koral} \citep{Sadowski+2013,Sadowski+2014}, ray-traced for snapshots over the time range $\sim 10^4-10^5\; GM/c^3$. Using these GRMHD snapshots, we perform GRRT using the code \texttt{ipole} \citep{Moscibrodzka&Gammie2018}, frequently utilized in EHT theoretical studies \citep{EHTC+2021b,EHTC+2022e,EHTC+2023,EHTC+2024c}.  This code evolves all four Stokes parameters in a general-relativistic framework, allowing us to compute images of both linear and circular polarization.  EHT has used a variety of polarized radiative transfer codes, including \texttt{BHOSS} \citep{Younsi+2012}, \texttt{grtrans} \citep{Dexter&Agol2009, Dexter2016}, \texttt{Odyssey} \citep{Pu+2016}, and \texttt{RAPTOR} \citep{Bronzwaer+2018, Bronzwaer+2020}. Together with \texttt{IPOLE}, all of them have been shown to produce consistent results \citep{Prather+2023}.

Because the mean free path of particles is much larger than the size of the system, ions and electrons are not believed to be in thermal equilibrium.  Consequently, the ion-to-electron temperature ratio as a function of plasma conditions is a major uncertainty \citep{Shapiro+1976,Rees+1982,Narayan&Yi1995b}.  In this work, electron temperatures are assigned in post-processing assuming the phenomenological $R-\beta$ prescription of \cite{Moscibrodzka+2016}.  That is,

\begin{equation}
    \frac{T_p}{T_e} = R_\text{high} \frac{\beta^2}{1+\beta^2} + R_\text{low} \frac{1}{1+\beta^2} 
\end{equation}
where $\beta = P_\text{gas} / P_\text{mag}$ and $P_\text{mag} = B^2/2$.  This prescription assigns an asymptotic temperature ratio of $R_\text{low}$ (often taken to be 1) to highly magnetized (low-$\beta$) regions, and a temperature ratio of $R_\mathrm{high}$ (usually varied between 1 and $\approx$160) to weakly magnetized (high-$\beta$) regions.  This is in qualitative agreement with simulations that evolve separate ion and electron temperatures explicitly, although this remains an active area of research \citep{Ressler+2015,Sadowski+2017,Ryan+2018,Chael+2019,Dihingia+2023}.
 
Since GRMHD simulations are scale free, we use the exact same GRMHD fluid snapshots to produce images for each of our objects.  For each source, the SMBH mass sets the length and time scales, the distance sets the angular size scale, and the 230 GHz flux density fixes  $\dot{M}$, the mass accretion rate of the fluid. Other quantities (magnetic field strength and internal energy) are rescaled appropriately. Following the procedure described in \citet{Qiu+2023}, we fit a slowly increasing\footnote{Slowly increasing $\mathcal{M}$ with time acts to counteract the draining and relaxation of the torus, which would otherwise result in a systematic decline in the flux density as the simulation proceeds.} $\mathcal{M}$, rescaling the density, internal energy, and magnetic field of the plasma to achieve the average 230 GHz flux density listed in \autoref{table:sources}.  Here, Eddington ratios are defined relative to the mass accretion rate that would be required to produce the Eddington luminosity for a fiducial radiative efficiency of 0.1.

Compared with other GRMHD model libraries \citep[e.g.,][]{EHTC+2019e}, for the main body of our paper, we significantly limit the scope of our parameter space to what we believe are representative values to enable us to model a variety of sources.  All models have dimensionless spin parameter $a_\bullet=0.9$, $R_\mathrm{high}=40$, $R_\mathrm{low}=1$, a magnetic field polarity aligned with the disk angular momentum, and only two different inclinations.  One set of images is computed with a viewing angle of $i=50^\circ$, which may be appropriate for the typical SMBH, while another set is computed with $i=160^\circ$, appropriate for \m87. A few other parameter combinations are discussed in \autoref{sec:other_parameters}, including $i=5\degree$, $a_\bullet = 0.9$, and $R_\mathrm{high}=160$.

In all of our runs, we included only MAD models, since they are preferred over their SANE counterparts by EHT studies of \sgra and \m87 \citep{EHTC+2021b,EHTC+2022e,EHTC+2023,EHTC+2024c}. However, we do note that the generality of this accretion state is an open area of research.  It will be interesting to see how well MAD models can represent the radio-quiet sources on our list, which have been proposed to be explained by weaker magnetic fields \citep[e.g.,][]{Sikora&Begelman2013}.  We expect that more weakly magnetized models will be even more Faraday thick and have stronger jet emission relative to their disk emission \citep[e.g.,][]{EHTC+2019e,EHTC+2021b}.  In such models, linear polarization structures like $\angle \beta_2$ would likely evolve even more strongly than we predict for MADs in \autoref{fig:tavg_vavg_beta2}.

The moderate value of $R_\mathrm{high}$ chosen for our models is broadly consistent with simulations which evolve ion and electron temperatures explicitly, we note that it is lower than the values preferred when comparing \sgra and \m87 models to EHT observations \citep{EHTC+2023,EHTC+2024c}. We confirm that $R_\mathrm{high}=160$ models are qualitatively similar in \autoref{sec:other_parameters}.  For larger values of $R_\mathrm{high}$ and $R_\mathrm{low}$, which would cool the model by construction, slightly larger accretion rates would be required to match the 230 GHz flux densities.  We do not expect qualitative changes in model images if different spin values are adopted, although the jet power will be strongly affected \citep[e.g.,][]{Tchekhovskoy+2011,Narayan+2022}.

We model only thermal electron distribution functions.  As a consequence, our models contain very little emission from their jets \citep{Davelaar+2019,Chatterjee+2021,Emami+2021,Cruz-Osorio+2022,Fromm+2022}, which are expected to be dominated by non-thermal electrons. As a spot-check, we refit $\mathcal{M}$ and inferred Eddington ratios with non-thermal $\kappa=5$ electron distribution functions instead of thermal, and found that that our Eddington ratios decreased to 37\% of their original values on average.

Finally, as is typically assumed in EHT studies, we neglect radiative transfer from cells with $\sigma > 1$, where GRMHD floors may inject artificial material into the simulations.  These assumptions are discussed in \autoref{sec:sum}. We note that ideal GRMHD simulations neglect radiative heating and cooling by construction.  While this is easily justifiable for very low Eddington ratio systems like \sgra and to a lesser extent \m87, the larger Eddington ratios inferred for these objects motivate additional radiative simulations \citep[e.g.,][]{Ryan+2018,Chael+2019,Yao+2021,Dihingia+2023}.

The limited set of models included here allows us to begin to forecast observational trends starting with a likely set of parameters identified in previous EHT studies.  However, we note that these trends will be model-dependent, and a much more extensive model selection process will be necessary to perform parameter estimation on the real data, as has been done for \m87 and \sgra \citep{EHTC+2019e,EHTC+2021b,EHTC+2022e,EHTC+2023,EHTC+2024c,EHTC+2025}.

\section{Results}

\subsection{Systematically Higher Eddington Ratios}
\label{sec:eddington_ratios}

\begin{figure*}[htb] 
\centering
\includegraphics[width=\textwidth]{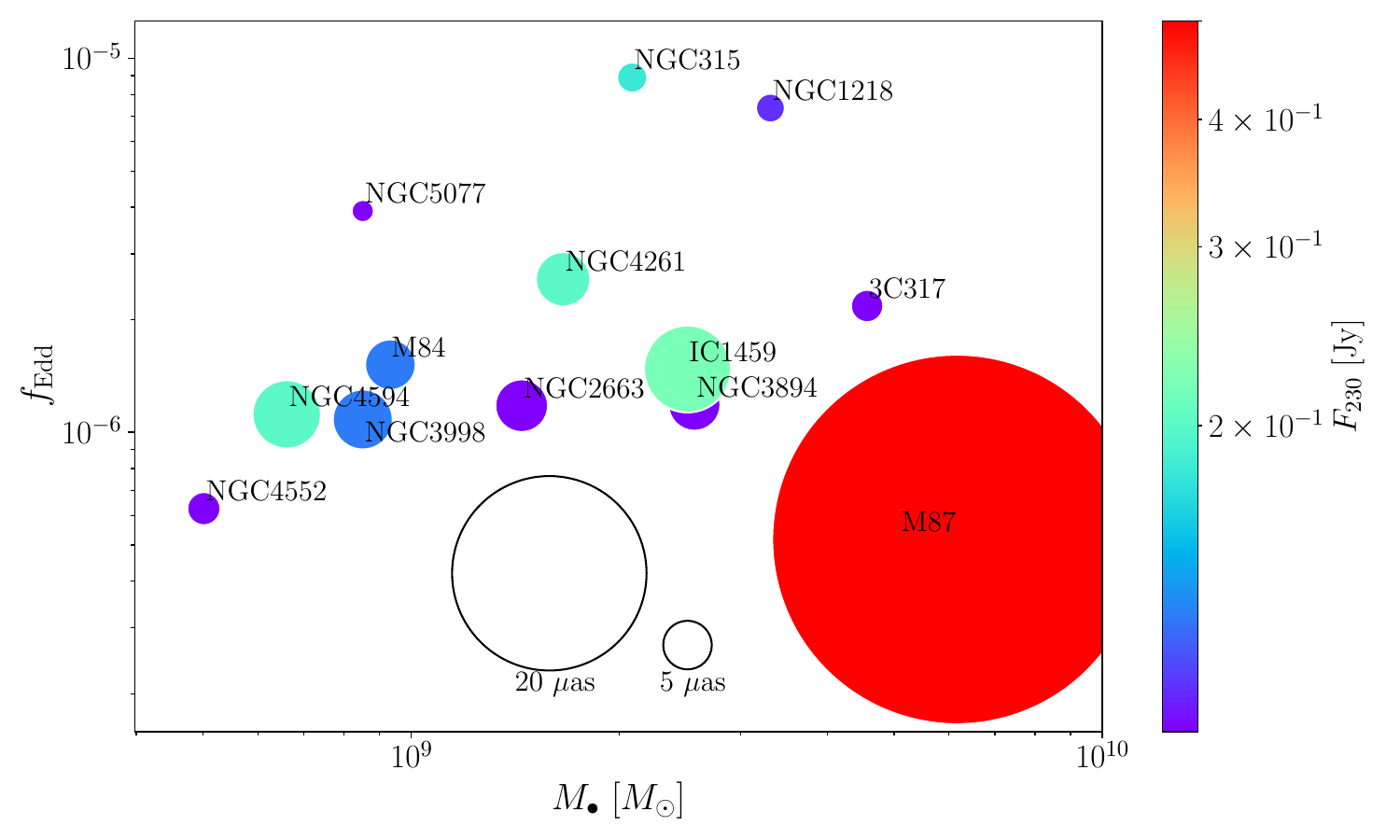}
\caption{Inferred Eddington ratio ($f_\mathrm{Edd}$) as a function of SMBH mass ($M_\bullet$) for our sources.  There is no strong correlation between the two properties.  In this figure, color encodes the 230 GHz flux density, size scales linearly with the angular diameter of these sources, and the upper left circles respectively show the EHT and BHEX beam sizes. All of these objects have slightly lower masses and higher Eddington ratios than \m87.}
\label{fig:sources}
\end{figure*}

Fitting for $\mathcal{M}$ for each source during ray-tracing gives us a model-dependent estimate of the Eddington ratio based on the 230 GHz flux density.  These are listed in \autoref{table:sources} and plotted as a function of $M_\bullet$ in \autoref{fig:sources}.  We predict larger Eddington ratios for all our sources than for both \m87 and \sgra, reaching up to $17$ times that of \m87 in the most extreme case of NGC 315$^*$.  This is a selection effect, since more distant sources need to have larger accretion rates to achieve a given flux density and become detectable.  

We caution that uncharacterized systematic errors dominate the error budget of our Eddington ratios on both the observational side (mass, distance, flux density, and compact flux fraction) and the theoretical side (MAD accretion, $R_\mathrm{high}$, $i$). An Eddington ratio estimate based solely on 230 GHz $\mathcal{M}$-fitting is likely only accurate to within an order of magnitude \citep[see e.g., Figure 13 of][]{EHTC+2021b}. Nevertheless, our findings suggest that extensions to the EHT will enable imaging of sources in a previously unexplored regime of Eddington ratio: up to $17$ and $254$ times that of $f_\mathrm{Edd} \sim 10^{-6}$ and $f_\mathrm{Edd} \sim 10^{-8}$ inferred for \m87 and \sgra respectively in the context of these models.  In this regime, radiative cooling effects (neglected in our simulations) will become more important, potentially lowering the temperature in the disk \citep[e.g.,][]{Ryan+2018,Chael+2019,Yao+2021,Dihingia+2023}.  

\begin{figure*}[htb]
\centering
\includegraphics[width=\textwidth]{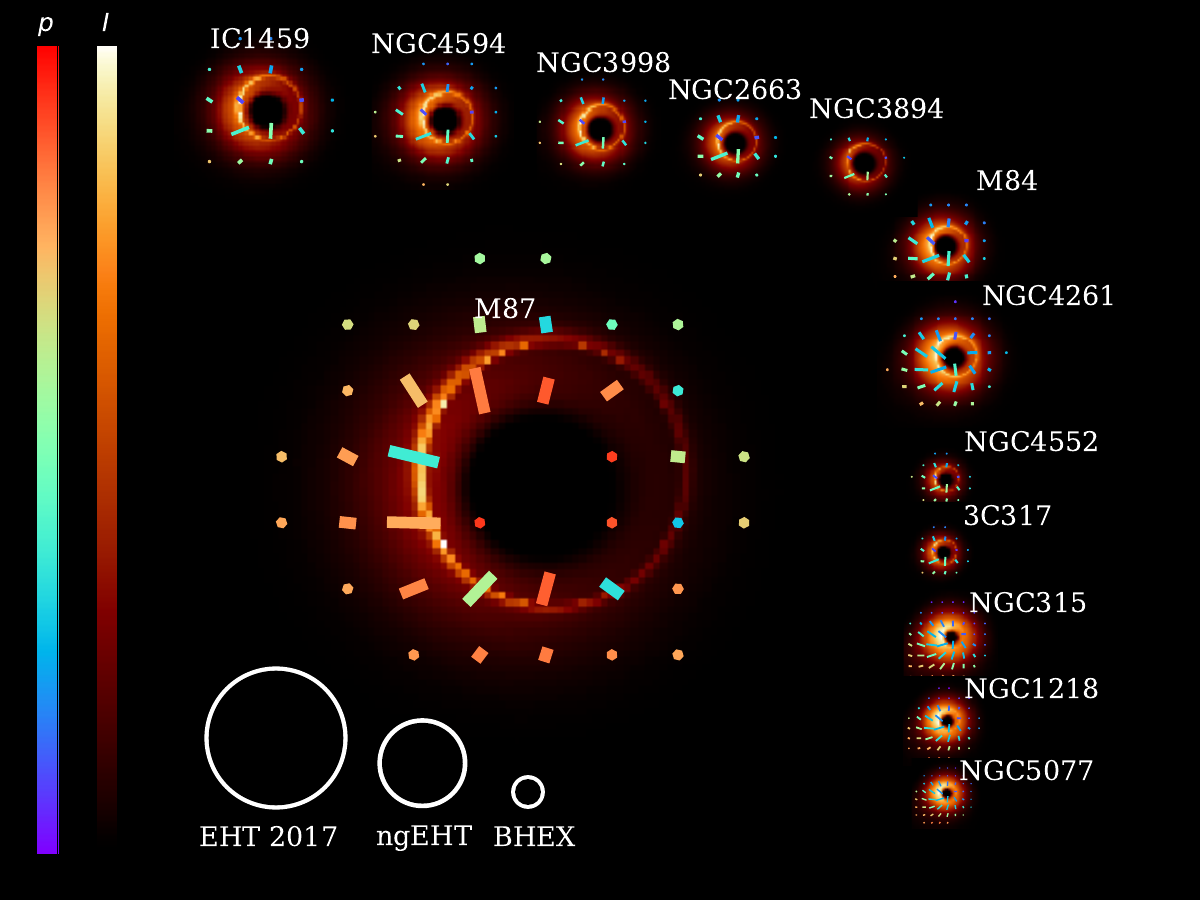}
\caption{Time-averaged total intensity and linear polarization images for each of our sources at 230 GHz, assuming a $160\degree$ viewing angle, in descending order of angular shadow size.  Images are plotted to scale, but the total intensity ($I$) color bar is scaled to each source individually.  Tick lengths scale with the total linear polarization in each image, while tick colors scale with the fractional linear polarization $p = P/I$, saturating at 70\%.  In the bottom left, we plot the beam sizes of several VLBI experiments: the current EHT 2017 array at 230\,GHz ($20\,\mu{\rm as}$), the ngEHT at 345\,GHz ($13\,\mu{\rm as}$), and BHEX ($5\,\mu{\rm as}$).}
\label{fig:gallery}
\end{figure*}

\begin{figure*}[htb]
\centering
\includegraphics[width=\textwidth]{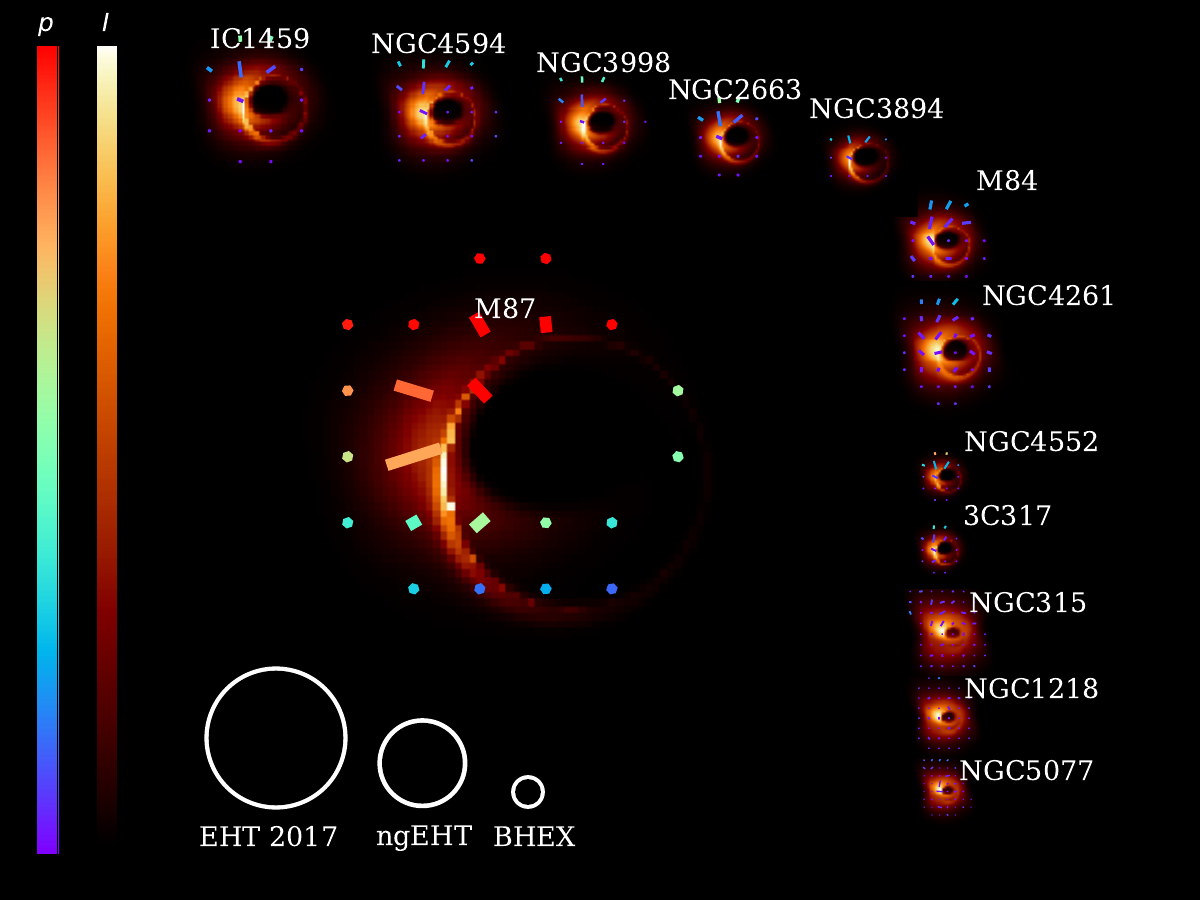}
\caption{As \autoref{fig:gallery}, but ray-traced at $50\degree$ inclination. At this larger inclination, model images are visibly much less polarized, the image is more asymmetric, and the model is more depolarized on the bottom, where emission passes through the Faraday thick disk midplane.}
\label{fig:gallery2}
\end{figure*}

We plot time-averaged polarized images for each simulation set of our sources at $i = 160\degree$ and $i=50 \degree$, in \autoref{fig:gallery} and \autoref{fig:gallery2} respectively. All SMBH angular sizes are plotted to scale and the total intensity maps are scaled individually to the maximum specific intensity in each image.  Recall, however, that sources may be up to an order of magnitude fainter than \m87 (\autoref{table:sources}).  Linear polarization is depicted with ticks whose lengths scale with the total amount of polarization ($P=\sqrt{Q^2+U^2}$) and whose colors scale with the fractional polarization ($p=P/I$) up to 70\%.  In the bottom left, we plot the beam sizes of several VLBI experiments: the current EHT 2017 array at 230\,GHz ($20\,\mu{\rm as}$) and BHEX ($5\,\mu{\rm as}$) for comparison.

All of these sources are expected to subtend significantly smaller angles than \m87, with shadow diameters in fact smaller than the nominal beam size of the present EHT.  This implies that super-resolution and modeling techniques will be crucial for inferring horizon scale structure of these sources \citep[e.g.,][]{Pesce+2022}.  By eye, our $i=160\degree$ images appear well-described by a polarized ring.  However, the $i=50\degree$ images may bear additional structures, such as a slash through middle of the ring, that cannot be captured by a ring model. More edge-on models appear to exhibit more source-dependent structural variation than more face-on models. Both the total intensity and polarization structures in these more edge-on images are more asymmetric \citep{Falcke+2000, Broderick+2009, psaltis+2015,medeiros+2022,Qiu+2023,EHTC+2024c}.  Some models (e.g., IC 1459$^*$), exhibit an offset between the total intensity and linearly polarized emission \citep[discussed in][and below]{Tsunetoe+2021,Tsunetoe+2022}.  Because of these inclination-based effects, independent jet-based inclination constraints will be very important for interpreting VLBI observations of these sources.

\begin{figure*}[htb]
\centering
\includegraphics[width=18cm]{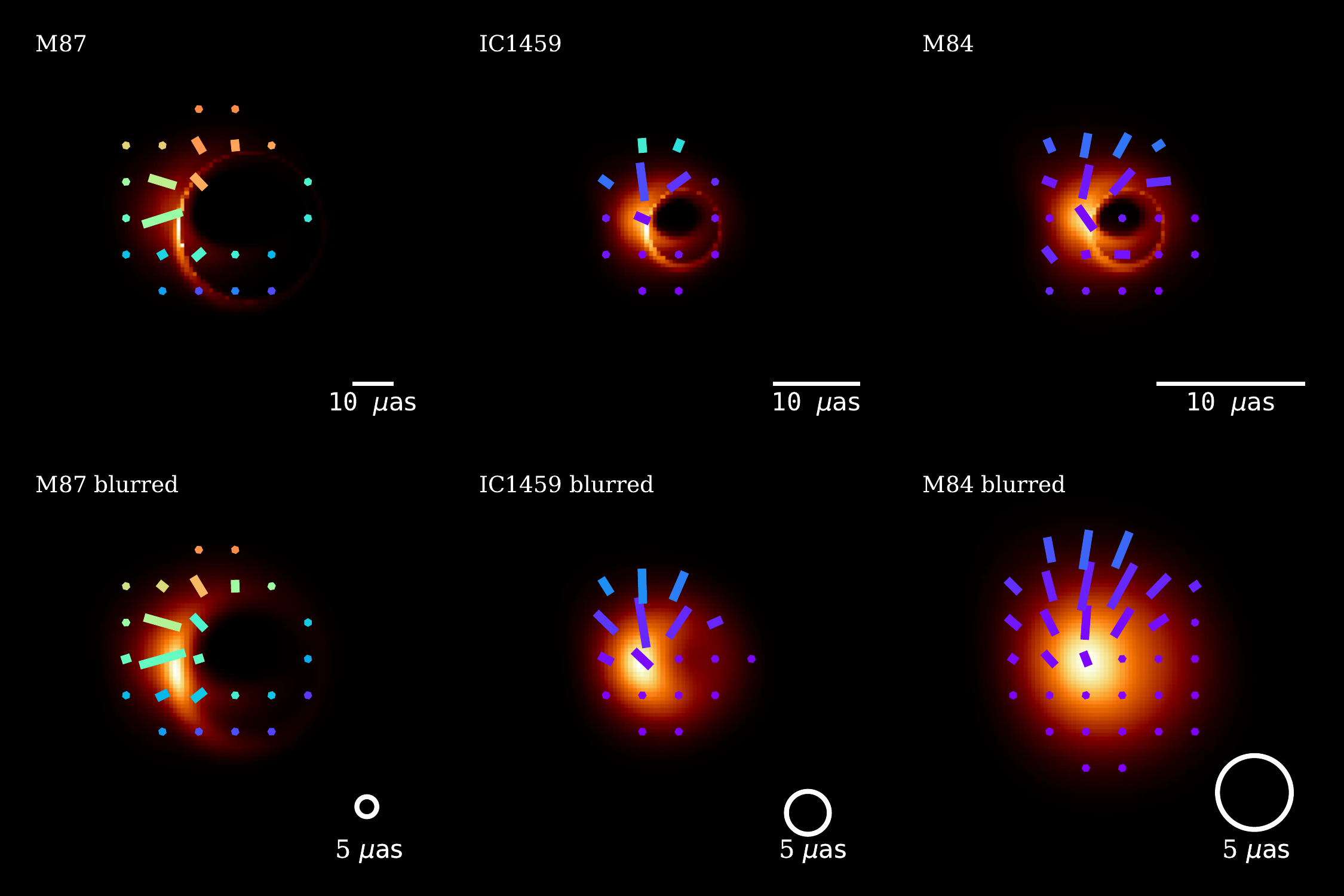}
\caption{To more clearly visualize the effects of Eddington ratio on our images, we plot time-averaged 230 GHz image of three different cases: M84$^*$ on the left ($f_\mathrm{Edd} = 1.51\times 10^{-6}$), IC1459 in the middle ($f_\mathrm{Edd} = 1.48\times 10^{-6}$), and M87* on the right ($f_\mathrm{Edd} = 5.2\times 10^{-7}$). The top row shows the time-averaged image, while the bottom row shows the images blurred to BHEX resolution (5$\mu as$). M87$^*$ is visibly much more optically thin and polarized.}
\label{fig:blurred}
\end{figure*}

The larger Eddington ratios in our sample explain the salient qualitative differences in image morphologies.  We highlight the differences between our models of \m87, IC1459$^*$, and M84$^*$ in \autoref{fig:blurred}, where each is plotted at $50\degree$ inclination. IC1459$^*$ and M84$^*$ has Eddington ratios approximately $3$ times that of \m87. Larger inferred Eddington ratios result in visibly more extended and optically thick emission for the models of our other sources than for those of \m87.  The bottom row of \autoref{fig:blurred} shows the photon rings blurred to BHEX resolution (5 $\mu as$), the photon ring being clearly visible for \m87 only, while the flux depression due to the ``inner shadow'' \citep{Chael+2021} is visible in all of them at perfect resolution. It is encouraging that these features remain accessible despite the moderate optical depths that we calculate from these images.

\begin{figure*}[htb]
\centering
\includegraphics[width=18cm]{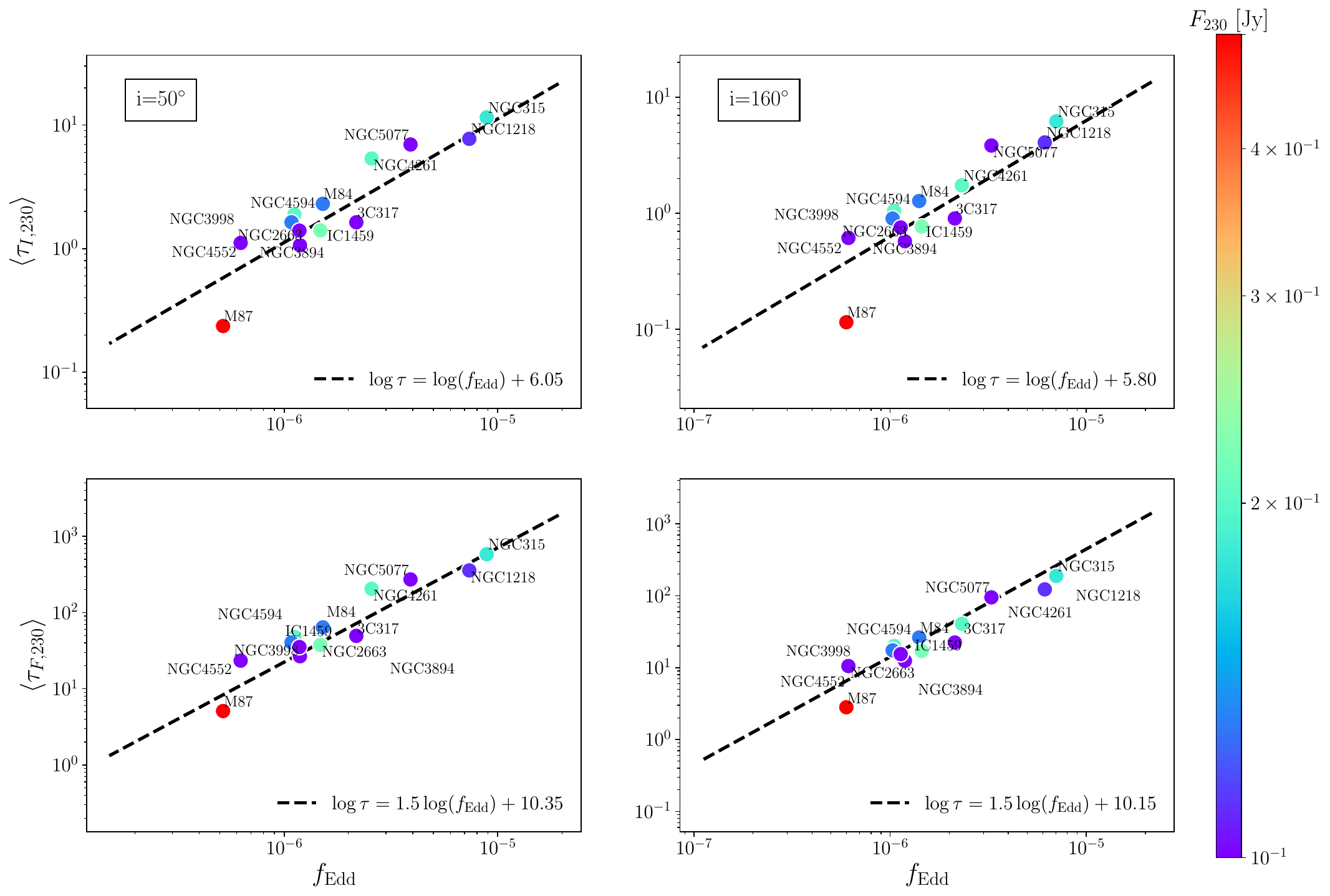}
\caption{Optical depth ($\langle\tau_\text{I,230}\rangle$) and Faraday rotation depth ($\langle\tau_\text{F, 230}\rangle$) as a function of Eddington ratio in our models $50\degree$ inclination (top) and $160 \degree$ inclination (bottom).  Due to their larger Eddington ratios, all sources are expected to have larger optical and Faraday depths than \m87, especially at less face-on viewing angles.}
\label{fig:tavg_favg}
\end{figure*}

Although these quantities are not directly observable, we compute characteristic optical and Faraday depths in order to understand the trends in our images. Since they vary by orders of magnitude across the image, and it is not obvious how to generically define a convenient location, we adopt an intensity-weighted average \citep[e.g.,][]{EHTC+2019e,EHTC+2021b}. For a single pixel, the optical depth $\tau_\text{I, 230}$ and Faraday depth $\tau_\text{F, 230}$ are each given by
\begin{equation}
\tau_\text{I, 230} = \int^\mathrm{observer}_\mathrm{source}\alpha_I ds
\end{equation}
and 
\begin{equation}
\tau_\text{F, 230} = \int^\mathrm{observer}_\mathrm{source}\rho_V ds.
\end{equation}
Here, $\alpha_I$ is the opacity, $\rho_V$ is the Faraday rotation coefficient \citep[e.g.,][]{Jones&Hardee1979}, and $s$ is the affine parameter describing the geodesic. Averaging over the entire image and weighting by the total intensity of each image pixel,
\begin{equation}
\langle\tau_\text{I, 230}\rangle =\frac{\int\tau_\mathrm{I, 230} I dxdy}{\int I dxdy}
\end{equation}
and similarly
\begin{equation}
\langle\tau_\text{F, 230}\rangle =\frac{\int\tau_\mathrm{F, 230} I dxdy}{\int I dxdy}
\end{equation}

We plot the image-averaged optical depth $\langle\tau_\text{I, 230}\rangle$ and Faraday depth $\langle\tau_\text{F, 230}\rangle$ of our models as a function of $f_\mathrm{Edd}$ in \autoref{fig:tavg_favg}. Values larger than unity are achieved by most models for both quantities.  In the upper row, we plot models $i=50\degree$ (more inclined) and in the lower row, we plot models $i=160\degree$ (less inclined).  

Internal Faraday rotation is the dominant mechanism by which our GRMHD models are scrambled and depolarized \citep{Moscibrodzka+2017,Jimenez-Rosales&Dexter2018,Ricarte+2020,EHTC+2021b}, and these results suggest that these effects may be more significant for this sample.  Both $\langle\tau_\text{I, 230}\rangle$ and $\langle\tau_\text{F, 230}\rangle$ are noticeably larger for the more inclined models.  In the case of $\langle\tau_\text{F, 230}\rangle$, when a model is more inclined, more of the colder and denser midplane intercepts our line-of-sight, increasing the Faraday rotation depth \citep{Ricarte+2020,EHTC+2024c}.

Our models roughly adhere to $\langle\tau_{I,230}\rangle \propto f_\mathrm{Edd}$ and $\langle\tau_{F,230} \rangle\propto f_\mathrm{Edd}^{3/2}$, plotted as dashed lines for reference.  We expect $\langle\tau_{F,230}\rangle \propto f_\mathrm{Edd}^{3/2}$ because the Faraday rotation coefficient $\rho_V \propto n_e B$ \citep{Gardner&Whiteoak1966}, $n_e$ is the electron number density and $B$ is the magnetic field strength.  In our GRMHD models, $f_\mathrm{Edd} \propto \mathcal{M}$, where $n_e \propto \mathcal{M}$, and $B \propto \sqrt{\mathcal{M}}$.  Therefore, $\langle\tau_{F,230}\rangle \propto \mathcal{M}^{3/2} \propto f_\mathrm{Edd}^{3/2}$.  For an isolated blob of emitting gas, one would expect $\langle\tau_{I,230}\rangle \propto \alpha_I \propto j_I \propto n_e B^2$ \citep[e.g.,][]{Leung+2011}, leading to $\langle\tau_{I,230}\rangle \propto f_\mathrm{Edd}^{2}$.  We find a shallower slope however, likely because increasing the Eddington ratio introduces new optically thin emission in addition to making previously emitting regions more optically thick. 

\subsection{Predictions for Polarized Structure}\label{sec:predictions}

\begin{figure*}[htb]
\centering
\includegraphics[width=18cm]{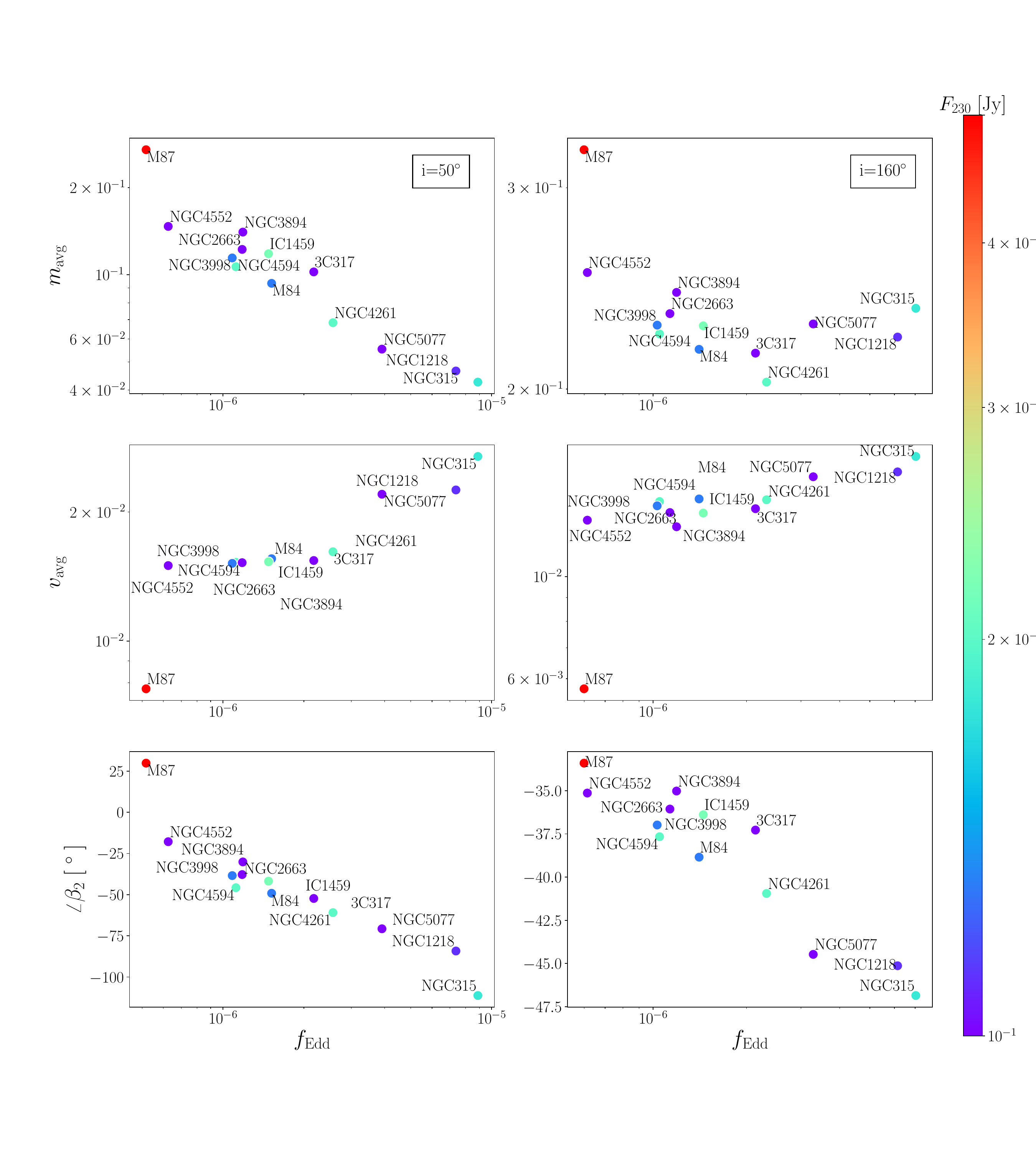}
\caption{Resolved linear polarization fraction ($m_\text{avg}$), resolved circular polarization fraction ($v_\text{avg}$), and pitch angle of linearly polarized morphology ($\angle\beta_2$) as a function of Eddington ratio ($f_\text{Edd}$), shown for $50\degree$ inclination in the left column and $160\degree$ inclination in the right column.  Since $\langle\tau_\text{I, 230}\rangle$ increases with $f_\mathrm{Edd}$ (\autoref{fig:tavg_favg}), for the more edge-on case of $i=50\degree$, $m_\text{avg}$ decreases with Eddington ratio.  Meanwhile, $v_\text{avg}$ increases with increasing $f_\text{Edd}$ in these models.  Despite our images originating from the same GRMHD snapshots, there is substantial evolution of $\angle \beta_2$ with $f_\mathrm{Edd}$ at $i=50\degree$, but much less evolution at $i=160\degree$.}
\label{fig:tavg_vavg_beta2}
\end{figure*}

Since $\langle\tau_\text{I, 230}\rangle$ and $\langle\tau_\text{F, 230}\rangle$ increase with Eddington ratio in our models, we anticipate quantitative changes in the polarimetric observables they produce. We compute $m_\text{avg}$ and $v_\text{avg}$, the image-average linear and circular polarization fractions on resolved scales, weighted by the total intensity of each pixel:
\begin{equation}
m_\text{avg} = \frac{\sum_i \sqrt{Q_i^2 +U_i^2}}{\sum_i I_i}
\end{equation}
\begin{equation}
v_\text{avg} = \frac{\sum_i |V_i/I_i|I_i}{\sum_i I_i}
\end{equation}
Note that these resolved values depend on the imaging resolution (beam size).  For our calculations, we report values where images are blurred with a Gaussian kernel with a FWHM equal to $20(M/M_{M87})(D_{M87}/D) \; \mu\mathrm{as}$, where $D$ is the distance, $M$ is the mass, and values with the subscript $M87$ correspond to M87 values.  This is done to ensure that these values are computed at comparable scales (in gravitational units) for all objects.

EVPA angles around the emission ring typically follows an azimuthal pattern, and we quantify this pattern following the approach of \cite{Palumbo+2020}. For an image with complex linear polarization $P(\rho,\varphi)=Q(\rho,\varphi)+iU(\rho,\varphi)$, the azimuthal modes have complex coefficients $\beta_m$ given by
\begin{equation}
\beta_m = \frac{1}{I_\mathrm{ann}} \int_{\rho_\text{min}} ^{\rho_\text{max}} \int^{2\pi}_0 P(\rho,\varphi) e^{-im\varphi}\rho d \varphi d\rho
\end{equation}
Here, $I_\text{ann}$ is the Stokes I flux density inside the annulus with radii $\rho_\text{min}$ and $\rho_\text{max}$. We set $\rho_\text{min}=0$  and $\rho_\text{max}$ large enough to cover the entire image. We are particularly interested in the $\beta_2$ coefficient since it gives an image-averaged measurement of EVPA rotational symmetry \citep{Palumbo+2020}.

We plot these quantities as a function of Eddington ratio in \autoref{fig:tavg_vavg_beta2}, for $i=50\degree$ in the left column and $i=160\degree$ in the right column.  As the Eddington ratio increases, larger optical and Faraday depths result in a downward trend in $m_\mathrm{net}$ at $50\degree$.  However, the linear polarization plateaus around $\sim 4.5$\% for $160\degree$.  We find that, especially for more face-on inclinations, much of the polarized emission comes from the forward-jet and travels through an evacuated funnel region, unaffected by the large Faraday depth behind it. Indeed, an additional check at the face-on case of $i=5\degree$ in \autoref{fig:appendix2}) in \autoref{sec:other_parameters} shows almost no evolution. For circular polarization, $v_\mathrm{avg}$ typically increases as a function of Eddington ratio. This may likely arise from increased Faraday conversion depths, shown to dominate the generation of Stokes V in models consistent with M$\,87^*$ \citep{EHTC+2023}. However, evolution in the intrinsic emission, Faraday rotation, and optical depth can also contribute \citep[e.g.,][]{Moscibrodzka+2021,Ricarte+2021,Joshi+2024}. 

We are particularly interested in the evolution of $\angle \beta_2$ as a function of $f_\mathrm{Edd}$.  Numerous studies of this quantity have demonstrated its sensitivity to the underlying magnetic field geometry, which in turn is affected by the SMBH spin via frame dragging \citep{Palumbo+2020,Ricarte+2022,Emami+2023,Qiu+2023,Chael+2023}. The frame dragging effect pulls the plasma around the SMBH, which in turn generates a spin-dependent magnetic field. As the spin increases, the magnetic field in the mid-plane wraps around the direction of the spin and becomes more toroidal. Since the synchrotron emission is linearly polarized perpendicular to the magnetic field, the magnetic field geometry could be measured by linear polarization ticks and thus $\angle \beta_2$. With our conventions, $\angle \beta_2 > 0^\circ$ implies counterclockwise rotation, consistent with the underlying accretion flow. For $i=50\degree$, we find a dramatic evolution of $\beta_2$ as a function of $f_\mathrm{Edd}$, even flipping sign.  That is, the sign of $\angle \beta_2$ does not simply encode the flow direction for substantially inclined sources.  The rapidity of this $\angle \beta_2$ evolution is striking, considering that the underlying GRMHD snapshots and temperature prescriptions are exactly identical: the same fluid has been only rescaled to different masses and Eddington ratios.  We expect that the exact evolution of $\angle \beta_2$ with $f_\mathrm{Edd}$ will be model-dependent, and difficult to predict (see e.g., our $a_\bullet=0$ models in \autoref{sec:other_parameters}).  This effect will complicate spin inference pipelines based on linear polarization, implying that different inference models may need to be developed for each source.

Revisiting \autoref{fig:gallery2} and \autoref{fig:tavg_favg}, the reasons for this effect can be understood.  Models with higher Eddington ratio have larger optical depth.  Thus, the emission region changes, and the polarization would therefore sample the magnetic field at a different location.  Models with higher Eddington ratio also have larger Faraday rotation depths, and this affects the image asymmetrically.  Polarization on the bottom of our image, which passes through the Faraday thick midplane, is suppressed.  For both of these reasons, the region producing the polarization is not necessarily the same region producing the total intensity.

This result motivates caution when interpreting $\angle \beta_2$ inferred from these sources, especially when lacking the spatial resolution to measure the offset of the linear polarization with respect to the total intensity.  On the other hand, for $i=160\degree$, the evolution of $\angle \beta_2$ with $f_\mathrm{Edd}$ is less severe, and additional checks for $i=5^\degree$ (as shown in \autoref{fig:appendix2}) in \autoref{sec:other_parameters} shows almost no evolution.  From \autoref{fig:gallery}, we can see that this is because the polarization continues to follow the total intensity, even at the largest Eddington ratios, due to the Faraday thin evacuated jet funnel.  This implies that $\angle \beta_2$ may be more easily interpretable for more face-on targets.  Note that the $\sim 30^\circ$ range in $\angle \beta_2$ spanned by our $i=160\degree$ models corresponds to an EVPA shift of $\sim 15^\circ$.  This is comparable to the precision with which $\angle \beta_2$ can even be measured in practice \citep{EHTC+2021a,EHTC+2024b}.  Fortunately, most of the sources in our sample have jets from which the inclination angle can in principle be independently constrained.

\section{Discussion and Conclusions} \label{sec:sum}

The EHT has successfully imaged two supermassive black holes, and extensions to the array including ngEHT and BHEX aim to measure horizon-scale image properties for dozens more. In this work, we introduced and explored 12 of the next-most-promising sources for millimeter VLBI observations based on their millimeter flux density and shadow size.  We performed GRRT to produce model images correctly scaled to the masses, distances, and 230 GHz flux densities of these objects, employing models with realistic physics.  We then used these simulated images to forecast population-level trends.  Our main results are as follows:

\begin{itemize}
    \item We predict that future targets will typically have Eddington ratios that are 3-4 times larger than \m87, probing accretion and jet launching in a new regime. As Eddington ratio increases, so too does optical and Faraday rotation depth.  As a result, we expect lower polarization fractions in higher Eddington ratio sources.
    \item We modeled two different inclinations, $i=160^\circ$ similar to \m87, and $i=50^\circ$ which may be more representative of the population as a whole.  Evolution in the image as a function of Eddington ratio is less pronounced for face-on viewing angles.
    \item For a fixed spin value, the morphology of the linear polarization (particularly $\angle \beta_2$) evolves strongly with Eddington ratio for $i=50^\circ$, but more weakly for $i=160^\circ$.  Inclination constraints will therefore be important for interpreting sparsely sampled and marginally resolved polarization structure for these sources, especially for the purpose of spin inference.
\end{itemize}

Despite the larger optical depths predicted by GRMHD, we find that we can recover source sizes by measuring the fall-off of the visibility amplitude as a function of (u,v)-distance with only a small systematic. Thus, mass measurements can be obtained for all sample sources from the ground with an accuracy that will typically exceed that of independent stellar or gas dynamical mass measurements \citep[see e.g.,][]{Simon+2024}.  For higher resolution imaging, which would be important if the underlying morphology is complex, extension of the EHT into space will be essential, provided such long baselines can achieve the necessary sensitivity. Further discussions on VLBI detectability could be found in \autoref{sec:vlbi}.

The heterogeneity of these targets will allow us to probe the parameter space related to accretion and jet launching in several new ways.  Future targets span a variety of host morphologies, and most, but not all, have known jets. This will allow us to study how jets are launched in a variety of environments.  Compared to \m87, these objects all have lower SMBH masses, lower galaxy stellar masses, and a variety of galaxy morphologies and environments.  The greater variety of Eddington ratios and inclinations than provided solely by \m87 and \sgra should enable us constrain the three-dimensional structure of the accretion disk as a function of Eddington ratio.  Of particular interest will be the horizon scale polarization of more highly inclined targets, since EHT studies prefer $i\sim 30^\circ$ for \sgra \citep{EHTC+2022e,EHTC+2024c}, and \m87 jet studies constrain $i \approx 17^\circ$ \citep{Walker+2018}.  At larger inclinations, models become more sensitive to Faraday rotation originating from otherwise invisible electrons, typically in the cold and dense mid-plane.  Models predict an asymmetry in the linearly polarized image owing to Faraday depolarization that should be tested observationally \citep{Ricarte+2020,Qiu+2023,EHTC+2024c}. 

While we highlight qualitative inferences based on GRMHD models, we caution that they are limited in numerous important ways. These are discussed in detail in \autoref{sec:GRMHD}, and we summarize them here:
\begin{itemize}
    \item {\bf Electron Thermodynamics}:  Our model lacks radiative heating and cooling by construction, and this may become a problem for targets with higher Eddington ratios. This motivatives additional radiative simulations. Our simulations also lacks non-thermal electron distributions and emission from $\sigma>1$ regions, which may reduce the jet component. 
    \item {\bf Compact Flux}: It is likely that the compact flux density from accretion is less than the flux density observed. However, since Eddington ratio evolves sublinearly with compact flux density, we believe that this problem will not significantly change our results. 
    \item {\bf Parameter Limitations}: Our model parameters are restricted to MAD simulations with $a_\bullet=0.9$ and $R_\mathrm{high}=40$, as well as two different inclinations $i=50\degree, 160\degree$. These represents typical cases from our simulation library, and three additional parameter combinations could be found in \autoref{sec:other_parameters}. 
\end{itemize}

\noindent Consequently, while the models here represent likely trends based on parameter combinations favorable from previous studies, much more extensive GRMHD library studies \citep[e.g.,][]{EHTC+2025} will remain important for inferring the parameters of each of these sources.
 
Although horizon scale structure will be accessible for each of these sources, they will for the most part only be modestly resolved, even for baselines to space.  This will make other sources of information important for interpreting VLBI data.  For example, \citet{Pesce+2022} used fits to geometric models of polarized rings, where mass and spin could be measured from the ring diameter and polarization structure respectively, to show that the ngEHT could realistically recover masses and spins for tens of sources from the ground. However, geometrical modelling of potential targets lack important realistic physics such as radiative transfer and optical depth, which are provided by GRMHD simulations. Our work reveals additional theoretical complications with respect to the inference of spin from the ring polarization structure in black hole images.  We find that for our $i=50\degree$ models, $\beta_2$ is not the dominant polarization mode due to optical and Faraday depth effects.  While mass inferences will not be affected by this, spin inferences may only be trustworthy for images where $|\beta_2| > |\beta_i|$ for all $i \neq 2$.  This likely necessitates $i \lesssim 40^\circ$ \citep[e.g.,][]{Qiu+2023}, which could be determined by e.g., jet inclination measurements.  It should also be tested if multi-frequency polarization measurements and resultant rotation measure maps can correct for this effect.

Observing the sources described in this paper, and more, will allow studies of black hole demographics over a more expansive sample of masses, Eddington ratios, inclinations, and spins than currently accessible to the EHT.  Imaging these sources should drive improvements in GRMHD simulations to better reproduce higher Eddington ratio sources, where radiative effects should become more important.  In future work, refined algorithms to estimate mass and spin will be tested on the more realistic (and optically and Faraday thick) models computed in this work, as well as simulations that include the missing physical processes described above.

\section{Acknowledgments} \label{sec:ack}

We thank George N. Wong for valuable feedback on the draft as well as Razieh Emami for useful discussions.  This project/publication is funded in part by the Gordon and Betty Moore Foundation (Grant \#8273.01). It is also made possible through the support of a grant from the John Templeton Foundation (Grant \#62286).  The opinions expressed in this publication are those of the author(s) and do not necessarily reflect the views of these Foundations.  This work was also supported by the National Science Foundation (AST-2307887, AST-1935980, and AST-2034306), the Brinson Foundation, and the Gordon and Betty Moore Foundation (GBMF10423). We acknowledge funding from ANID Chile via Nucleo Milenio TITANs (NCN2023$\_$002), Fondecyt Regular (1221421) and Basal (FB210003).

This research has made use of data obtained from the Chandra Source Catalog, provided by the Chandra X-ray Center (CXC) as part of the Chandra Data Archive. This research has made use of the NASA/IPAC Extragalactic Database (NED), which is funded by the National Aeronautics and Space Administration and operated by the California Institute of Technology.

\vspace{5mm}
\software{AstroPy \citep{Astropy+2013,Astropy+2018}, Matplotlib \citep{Hunter:2007}, 
NumPy \citep{harris2020array},
KORAL \citep{Sadowski+2013,Sadowski+2014},
IPOLE \citep{Moscibrodzka&Gammie2018},
\texttt{eht-imaging} \citep{Chael+2022}}.

\appendix
\section{Properties of Potential Targets}\label{appendix:properties}

\begin{itemize}

\item \sgra is the supermassive black hole located at the center of the Milky Way. One of the most studied sources, it has a distance of $0.008 \pm 0.00004$ Mpc and a mass of log($M_\bullet$ [M$_\odot$]) = 6.6$^{+0.0014}_{-0.0014}$ from stellar dynamics \citep{Ghez+2008, Gillessen+2009, Gillessen+2017, Do+2019, GravityCollab+2018, GravityCollab+2019}. Its 230 GHz flux density is $2.4 \pm 0.28$ Jy \citep{EHTC+2022c}, and the stellar mass of its host galaxy is $10^{10.8}$ M$_\odot$ \citep{Licquia+2015}. It has been studied extensively by the EHT \citep{EHTC+2022a, EHTC+2022b, EHTC+2022c, EHTC+2022d, EHTC+2022e, EHTC+2022f, EHTC+2024b, EHTC+2024c}.

\item M87 is an elliptical galaxy with a distance of $17 \pm 0.62$ Mpc \citep{Blakeslee+2009, Bird+2010, Cantiello+2018, EHTC+2019d} and a mass of log($M_\bullet$ [M$_\odot$]) = 9.8$^{+0.026}_{-0.027}$ from stellar dynamics \citep{Gebhardt+2011, Gebhardt&Thomas2009, Walsh+2013, Macchetto+1997}.  Its 230 GHz flux density is $0.64 ^{+0.39}_{-0.08}$ Jy \citep{EHTC+2019d} \footnote{We have adopted a value of 0.5 Jy for ray-tracing to maintain consistency with other GRMHD libraries computed for the EHT, facilitating easier comparison.}. It has been studied extensively by the EHT \citep{EHTC+2019a,EHTC+2019b,EHTC+2019c,EHTC+2019d,EHTC+2019e,EHTC+2019f,EHTC+2021a,EHTC+2021b,EHTC+2023,EHTC+2024a}, and has a well-studied jet that has been observed \citep[e.g.][]{Mertens+2016, Lu+2023, Kim+2024} and now studied at VLBI frequencies \citep[e.g.][]{Walker+2018}. A $17.2\pm 3.3\degree$ viewing angle has been estimated from jet-to-counterjet brightness and velocity ratios \citep{Mertens+2016}.

\item IC 1459 is an elliptical LINER \citep[e.g.][]{Sadler1987, Philips+1986} galaxy located in a loose galaxy group with 8 bright members \citep{Brough+2006}. It is a slowly rotating galaxy with a counter-rotating core \citep{Franx&Illingworth1988, Prichard+2019}. It has a distance of $29\pm 3.7$ Mpc and a mass of log($M_\bullet$ [M$_\odot$]) = 9.4$^{+0.077}_{-0.035}$ from stellar dynamics \citep{gultekin+2019}. Its 230 GHz flux density, measured at 0\farcs9 resolution with ALMA, is $0.22 \pm 0.021$ Jy \citep{Ruffa2019}. It has symmetric jets \citep{Tingay+2015}. 

\item NGC 4594 (also known as M104, the Sombrero Galaxy) is a LINER \citep{Kormendy1988} galaxy hosting a LLAGN. It is a well-known potential EHT target \citep{Bandyopadhyay+2019, Fish+2020}, and also the closest galaxy on this list with a distance of $9.9 \pm 0.82$ Mpc. Its mass is log($M_\bullet$ [M$_\odot$]) = 8.8$^{+0.025}_{-0.027}$ from stellar dynamics \citep{gultekin+2019}. Its 230 GHz flux density measured at arcsec scales, from the ALMA Calibrator Source Catalogue \footnote{https://almascience.nrao.edu/sc/}, is $0.20 \pm 0.01$ Jy. It is a well-studied AGN with small-scale radio jets \citep{Hada+2013}. From the jet-to-counterjet brightness ratio, the jet viewing angle is estimated at $66\degree^{+4\degree}_{-6\degree}$ \citep{Yan+2024}.  

\item NGC 3998 is a lenticular Seyfert 1 \citep[e.g.][]{Knapp+1985, Reichert+1992, Devereux+2011}/ LINER \citep[e.g.][]{Heckman1980, Taylor+1998} galaxy hosting a LLAGN, located in the outer areas of the Ursa Major group. Kinematical observations show that NGC 3998 is tidally stripped of dark matter \citep{Boardman+2016}. It has a distance of $14 \pm 1.3$ Mpc and a mass of log($M_\bullet$ [M$_\odot$]) = 8.9$^{+0.035}_{-0.035}$ from stellar dynamics \citep{gultekin+2019}. New VLBA 43 GHz imaging (Ramakrishnan et al., in prep.) reveals a flux of $0.13 \pm 0.024$ Jy at 0.5 mas resolution; given its relatively flat spectrum at mas-scales, we use this value for the 230 GHz flux. VLBI observations have identified a jet-like structure on the northern side of its nucleus \citep{Filho+2002, Helmboldt+2007}, and further observations revealed a kpc-size one-sided jet \citep{Frank+2016}. A $10\degree-21\degree$ jet viewing angle range has been estimated \citep{Principe+2020}.

\item NGC 4261 is an elliptical FR I LINER \citep[e.g.][]{Jaffe+1996} galaxy in the Virgo cluster, hosting a LLAGN. It has been targeted numerous times over the years, with a famously resolved disk observed with HST in 1993 \citep{Jaffe+1993}. It has a distance of $31 \pm 1.6$ Mpc and a mass of log($M_\bullet$ [M$_\odot$]) = 9.2$^{+0.11}_{-0.099}$ from CO measurements \citep{Boizelle+2021}. Its 230 GHz flux at 0\farcs2 resolution is $\sim$0.2-0.25 mJy \citep{Boizelle+2021}. It has a two-sided jet \citep{Haga+2015,Yan+2023}, and a nuclear disk of dust roughly perpendicular to the radio jet \citep{Jaffe+1993}. A $63 \pm 3 \degree$ viewing angle has been estimated from jet-to-counterjet brightness and apparent jet speed \citep{Piner+2001}. Its relatively compact cool core could be classified as a galactic corona \citep{OSullivan+2011}, a cool core with radii on the scale of a few kiloparsecs. 

\item NGC 2663 is an elliptical galaxy with a distance of $28 \pm 1.4$ Mpc and a mass of log($M_\bullet$ [M$_\odot$]) = 9.2$^{+0.58}_{-0.61}$, determined from M-$\sigma$ relationship of \cite{Saglia+2016}  and the $\sigma$ in \cite{Gultekin+2011}. Although its SED is poorly sampled at present, SED modeling suggests that it may be a viable source for resolving event horizon structure (Nagar et al. in prep.).  Lacking millimeter observations, we adopt its mas-scale 8.4 GHz flux density of $0.084 \pm 0.0084$ Jy from the VLBA Calibrator Catalogue, but note that this may be a significant underestimate. It features some of the longest collimated jets known, extending to 355 kpc from one side to the other \citep{Velovic+2022}.

\item NGC 3894 is an elliptical LINER \citep{Goncalves&Roos2004} galaxy hosting a LLAGN \citep{Balasubramaniam+2021}, with a distance of $50 \pm 2.5$ Mpc and a mass of log($M_\bullet$ [M$_\odot$]) = 9.4$^{+0.48}_{-0.48}$, determined from M-$\sigma$ relationship of \cite{Saglia+2016}  and the $\sigma$ in \cite{vandenbosch2016}. Its 0\farcs5 mas resolution 43 GHz flux is 60 mJy (Ramakrishnan et al., in prep.). We adopt its 230 GHz $0.058 \pm 0.015$ Jy flux density measured by SCUBA \citep{Anton2004}. It has symmetric relativistic parsec-scale jets \citep{Taylor+1998}; from VLBA observations between 5 and 15 GHz, a viewing angle of $10\degree < i < 21\degree$ is estimated \citep{Principe+2020}.

\item M84 (NGC 4374) is an elliptical or lenticular galaxy in the Virgo cluster. It is one of few known systems close enough for resolved Chandra observations in the Bondi radius \citep{Bambic+2023}. M84 has depressions in emissivity in the Northern and Southern regions of the AGN \citep{Bambic+2023}, characteristic of the radio lobes associated with FR I \citep{Fanaroff+Riley1974} radio jet activity \citep{Laing+Bridle1987}. It has a distance of $19 \pm 0.60$ Mpc and a mass of log($M_\bullet$ [M$_\odot$]) = 9.0$^{+0.042}_{-0.043}$ from gas measurements \citep{gultekin+2019}. At 230 GHz, linear polarization was not detected by the SMA, suggesting significant Faraday depolarization \citep{Bower+2017}.  Its 230 GHz flux density was measured at $0.13 \pm 0.015$ Jy at 0\farcs9 resolution. Interestingly, it is highly variable (at $\sim$5\arcsec\ scales), in the range $\sim$0.080-0.225 Jy \citep{Chen2023}. It has a compact core and a single jet at pc-scales \citep{Nagar+2002}, and double FR I type jets at kpc scales \citep{Laing+Bridle1987}. Two different jet viewing angles have been estimated: \cite{Meyer+2018} determined an angle of $74\degree^{+9\degree}_{-18\degree}$ based on the apparent speed and jet-to-counterjet flux ratio of the outer jet on a scale of hundreds of parsecs, while \cite{Wang+2022} estimated an angle of $58\degree^{+17\degree}_{-18\degree}$ using the proper motion and the jet-to-counterjet flux ratio from VLBI observations of the inner jet, which we have adopted.

\item NGC 4552 (M89) is an elliptical LINER \citep[e.g.][]{ho+1997, Gonzalez-Martin2009} galaxy, with a distance of $15 \pm 0.99$ Mpc and a mass of log($M_\bullet$ [M$_\odot$]) = 8.7$^{+0.051}_{-0.051}$ from stellar dynamics \citep{Saglia+2016}. Its 230 GHz flux observed at $\sim$5\arcsec\ scales by SMA is highly variable: $\sim$18--38 mJy \citep{Chen2023}, and we adopt the median value of $0.027 \pm 0.009$ Jy. It has a compact core and twin jets at pc-scales \citep{Nagar+2002}. Although optical observations identified a jet extending 10 arcminutes \citep{Malin1979}, later observations concluded that the apparent jet may instead be a tidal tail from a galaxy interaction \citep{Clark+1987, Katsiyannis+1998}. In recent years, it was observed that M89 is radio-dim \citep{Wojtowicz+2023}, raising the possibility of the absence of a jet.

\item 3C 317 is a borderline Seyfert 2/LINER \citep{Asmus+2015} cD galaxy hosting a LLAGN in the X-ray cooling flow cluster Abell 2052, offering the opportunity to study jet launching in an environment where AGN feedback is less effective than in M87 \citep{Zhao+1993}. It has a distance of $140 \pm 7.2$ Mpc and a mass of log($M_\bullet$ [M$_\odot$])= 9.7$^{+0.022}_{-0.032}$ from M-$L_\mathrm{bulge}$ measurements \citep{Mezcua+2018}.  The ALMA Calibrator Source Catalogue lists 230 GHz flux densities of 0.032 to 0.034 Jy from 2017 to 2019, and we adopt the value of $0.034 \pm 0.002$ Jy. Unlike the typical FRI galaxies, which are characterized by cores, twin jets and lobes, 3C 317 has an amorphous halo around its bright core \citep{Venturi+2004}. It has two opposing jets with twisted morphology \citep{Venturi+2004}. 

\item NGC 315 is a giant elliptical FR II galaxy hosting an LLAGN \citep{Tomar+2021}. Located in the Zwicky cluster, it has a distance of $70 \pm 3.5$ Mpc and a mass of log($M_\bullet$ [M$_\odot$])= 9.3$^{+0.064}_{-0.033}$ from CO measurements \citep{Boizelle+2021}. We adopt the 230 GHz flux density of $0.18 \pm 0.009$ Jy from the ALMA Calibrator Source Catalogue; multiband VLBI imaging data are also available \citep{Park+2021}. The spectral energy distribution (SED) for this AGN has been measured with a wavelength range from radio to X-ray, and a bolometric luminosity of Lbol$\sim 1.9\times 10^{43}$ erg s$^{-1}$ and a corresponding Eddington ratio L/L$_{Edd}$ of $4.97\times 10^{-4}$ have been obtained \citep{Gu+2007}, similar to our inferred Eddington ratio of $1.7\times 10^{-4}$ from the millimeter flux alone. Its jet has been observed numerous times over the years. 

\item NGC 1218 (3C 78) is a FR I broadline Seyfert 1 \citep{Waddell&Gallo2020}/BL Lac \citep{Nemmen+2010} radio galaxy, with a distance of $120 \pm 5.9$ Mpc and a mass of log($M_\bullet$ [M$_\odot$]) = 9.5$^{+0.48}_{-0.48}$ from the M-$\sigma$ relationship of \cite{Saglia+2016} and the $\sigma$ in \cite{vandenbosch2016}. The 230 GHz flux density at $\sim$5\arcsec\ resolution as measured by the IRAM 30m telescope is $0.11 \pm 0.01$ Jy \citep{Agudo2014} (which we adopt in this work), and the $\sim$23\arcsec\ resolution 350 GHz flux is 0.278 Jy \citep{Quillen2003}. A one-sided $\sim$ arcsecond long jet, similar to that of M87 (although smaller in projected length) has been observed at radio and optical frequencies \citep[e.g.][]{Trussoni+1999, Sparks+1995, Saikia+1986}. The jet profile is approximately conical on pc to kpc scales, and three compact knots, in addition to the core, were observed \citep{Roychowdhury+2024}. It is one of the three galaxies in this sample with high inferred Eddington ratios (i.e on the order of $10^{-4}$), L/L$_{Edd}=13.3\times 10^{-4}$.

\item NGC 5077 is an lenticular galaxy with LLAGN and LINER \citep{deFrancesco+2008} characteristics, the brightest of a small group of 8 galaxies \citep{Sanchez-Portal+2004,Tal+2009}. It has a distance of $39 \pm 8.4$ Mpc and a mass of log($M_\bullet$ [M$_\odot$])= 8.9$^{+0.18}_{-0.32}$ from gas measurements \citep{gultekin+2019}. Featuring a stellar core counter-rotating with respect to its main stellar body, properties of its misaligned gas suggests a past gas-rich merger \citep{Raimundo2021}. On arcsec-scales it hosts a flat spectrum radio core at frequencies between 1.4 and 15 GHz \citep{Nagar+1999}. Its 230 GHz flux density is variable, with values of $\sim$0.060-0.20 mJy at $\sim$5\arcsec\ resolution over the 2015-2019 period \citep{Chen2023}. We adopt the flux of $0.068 \pm 0.007$ Jy measured at 285 GHz listed in the ALMA Calibrator Source Catalogue.

\end{itemize}

\section{Testing Additional Parameters}
\label{sec:other_parameters}
\begin{figure*}[htb]
\centering
\includegraphics[width=\textwidth]{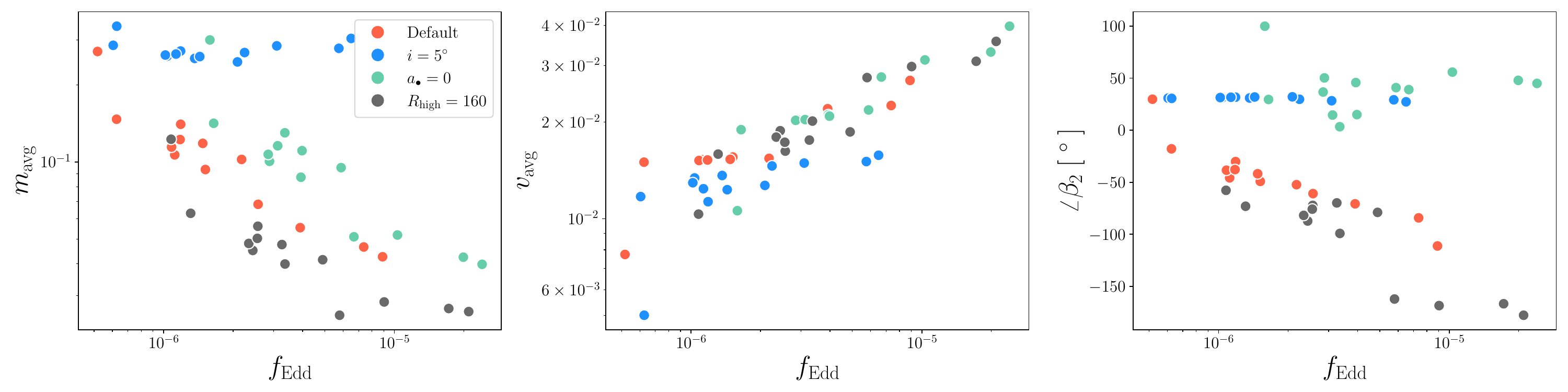}
\caption{Resolved linear polarization fraction ($m_\text{avg}$), resolved circular polarization fraction ($v_\text{avg}$), and pitch angle of linearly polarized morphology ($\angle\beta_2$) as a function of Eddington ratio ($f_\text{Edd}$). The default parameters are $i=50\degree$, $a_\bullet = 0.9$, and $r_\mathrm{high}=40$, as shown in red. A single parameter is changed for each of the other runs: $i=5\degree$ is shown in blue, $a_\bullet = 0.9$ is shown in green, and $R_\mathrm{high}=160$ is shown in grey. For all runs except $i=5\degree$, $m_\text{avg}$ decreases with Eddington ratio, while in every case, $v_\text{avg}$ consistently decreases with Eddington ratio. The evolution of $\angle \beta_2$ with $f_\mathrm{Edd}$ is minimal at $i=5\degree$, with greater evolution for all other runs.}
\label{fig:appendix2}
\end{figure*}

In the main body of the paper, the parameter space explored is limited to $a_\bullet=0.9$, $R_\mathrm{high}=40$, $R_\mathrm{low}=1$, and two different inclinations at $i=50^\circ$ and $i=160^\circ$. Here, we spot check the behavior of a few other relevant parameter combinations. These runs feature only MAD models, and the default parameters are $i=50\degree$, $a_\bullet = 0.9$, and $R_\mathrm{high}=40$. In each of the other runs, a single parameter is replaced, namely: $i=5\degree$, $a_\bullet = 0.9$, and $R_\mathrm{high}=160$, as seen in \autoref{fig:appendix2}. The results produced from these runs are consistent with the typical cases in the main paper, motivating comparable discussions and conclusions.

Similar to \autoref{fig:tavg_vavg_beta2}, for these runs, we plot $m_\text{avg}$, $v_\text{avg}$, and $\angle\beta_2$ as a function of Eddington ratio ($f_\text{Edd}$). 
For linear polarization, $m_\mathrm{avg}$ decreases with increasing Eddington ratio in all runs, except for $i=5^\degree$ as expected, due to the increased optical and Faraday depth. The plateau observed at $i=5^\degree$ resembles the behavior seen at $i=160^\degree$, which can be attributed to the face-on inclination, as discussed in \autoref{sec:predictions}. For circular polarization, $v_\mathrm{avg}$ exhibits a consistent upward trend with Eddington ratio among models.

For $\angle \beta_2$ as a function of $f_\mathrm{Edd}$, there is almost no evolution at $i=5\degree$, and all of the $\angle \beta_2$ has positive sign (consistent with the underlying accretion flow), as expected for a face-on target due to the Faraday thin evacuated jet funnel. Our $R_\mathrm{high}=160$ models behave similarly to our $R_\mathrm{high}=40$ models, except that the $\angle \beta_2$ values are lower. This is because with higher $R_\mathrm{high}$, the Eddington ratio is higher, which means that there is more cooling and thus more Faraday rotation.  The $a_\bullet=0$ models exhibit a weaker but non-monotonic trend, underscoring that the exact evolution of $\angle \beta_2$ is likely model-dependent and difficult to predict a priori.

\section{Detectability with VLBI}
\label{sec:vlbi}

\begin{figure*}[htb]
\centering
\includegraphics[width=18cm]{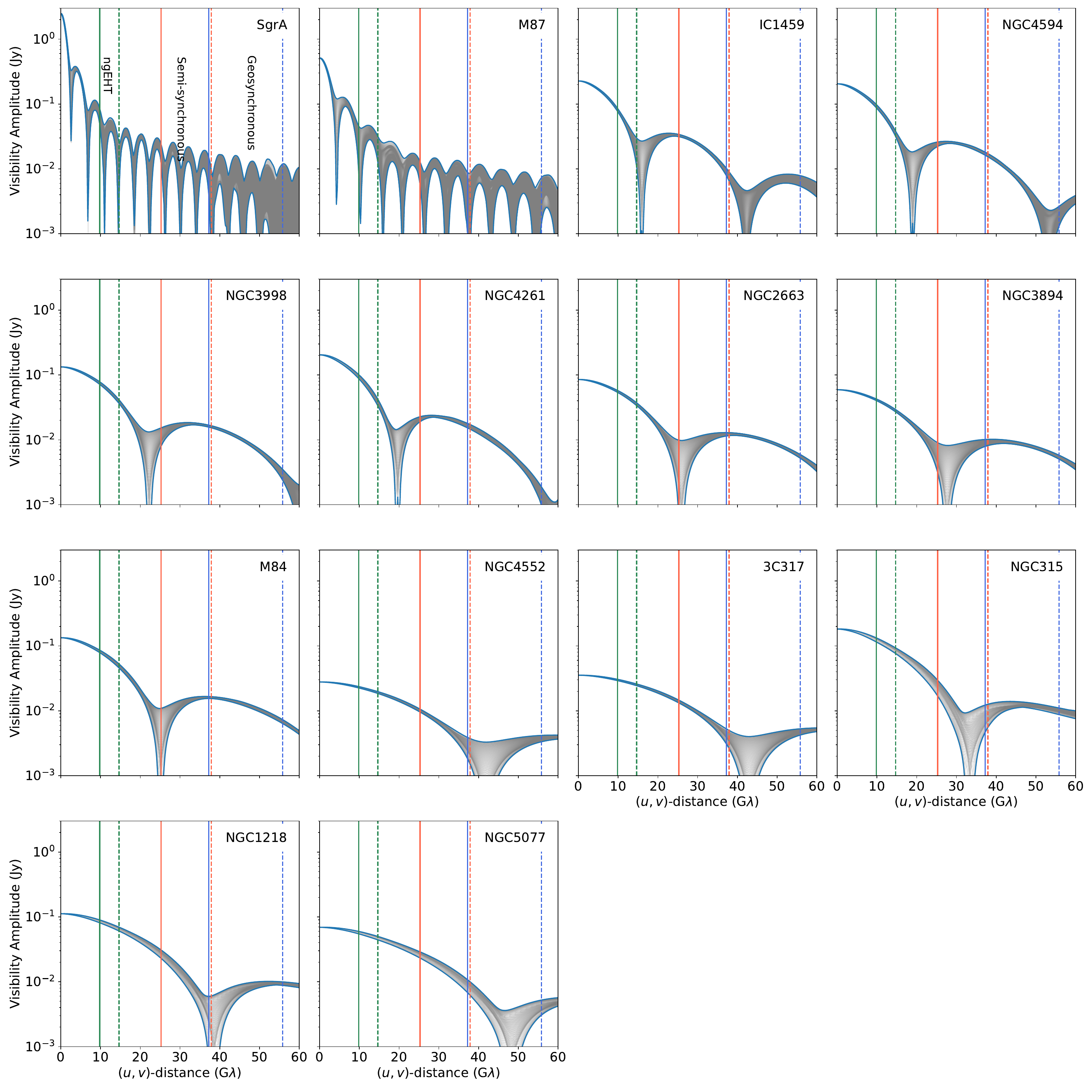}
\caption{Visibility amplitudes as a function of (u,v) distance derived from the time-averaged 230 GHz images of our sources at $160\degree$, with minimum and maximum values bounded by light blue lines.  We demarcate the nominal resolution of EHT/ngEHT in green, an array enhanced with an orbiter in semi-synchronous orbit in red, and an array enhanced with an orbiter in geosynchronous orbit in blue, where solid lines correspond to 230 GHz and dashed lines correspond to 345 GHz.  A ground-based array can constrain source sizes of these objects, but extensions into space are necessary to access the first null. We note that these visibility amplitudes do not represent any average or single observation run, and only indicate a ``typical'' observation.}
\label{fig:visibility_amplitude}
\end{figure*}

At present, VLBI observations of \m87 and \sgra at 230 GHz contain baselines with a maximum (u,v) distance of approximately 8 G$\lambda$, with a sensitivity of approximately $10^{-3}$ Jy on only the most sensitive baselines \citep{EHTC+2024a}.  By increasing the number of stations, increasing their recording bandwidths, and increasing the frequency of observation to 345 GHz, continued developments of the ground-based array are expected to achieve 15 G$\lambda (\approx D_\oplus / 0.87 \ \mathrm{mm}$ where $D_\oplus$ is the diameter of the Earth) \citep{Johnson+2023}.  To reach yet higher resolution, extensions of the array into space are necessary to achieve longer baselines.  BHEX aims to augment the ground-based array with a single orbiter in semi-synchronous orbit \citep{Johnson+2024}.  Additional stations in space are being considered to further improve the (u,v)-coverage. \citep{Kudriashov+2021,Roelofs+2021,Shlentsova+2024}. Currently, the BHEX mission is planned to go observe between 86 GHz and 345 GHz, while previous studies have considered frequencies all the way up to 690 GHz \citep[e.g.][]{Kudriashov+2021, Roelofs+2021}. 

Using our 345 GHz image models at $50\degree$, we assess the capability of extensions of the array to measure properties of our sources.  By Fourier transforming our time-averaged images, we visualize the visibility amplitudes of each source in \autoref{fig:visibility_amplitude}. The vertical lines in both images mark the maximum (u,v) distance accessible to EHT/ngEHT, an array including a semi-synchronous satellite, and an array including a geosynchronous satellite.  Scales accessible at 230 GHz and 345 GHz are shown with solid and dashed lines respectively.  Note that these lines demarcate only spatial resolution without considering sensitivity requirements, which may be the limiting factor for observing some of these sources with ground-space baselines. While a typical night of observation would have more turbulent structures, these visibility amplitudes can be used to set expectations for baseline sensitivity requirements.

For none of our new sources is the first minimum directly accessible from the ground.  This underscores the necessity of model-fitting for studying SMBH demographics with the ngEHT \citep[e.g.][]{Pesce+2021,Pesce+2022,Broderick+2011,Broderick+2009, Fish+2009a, Fish+2009b}.  Assuming sufficient baseline sensitivity, the addition of a semi-synchronous satellite operating at 345 GHz would allow direct access to the first null for 8 of these sources, while a geosynchronous satellite allows access to all 12. 

\begin{figure*}[htb]
\centering
\includegraphics[width=18cm]{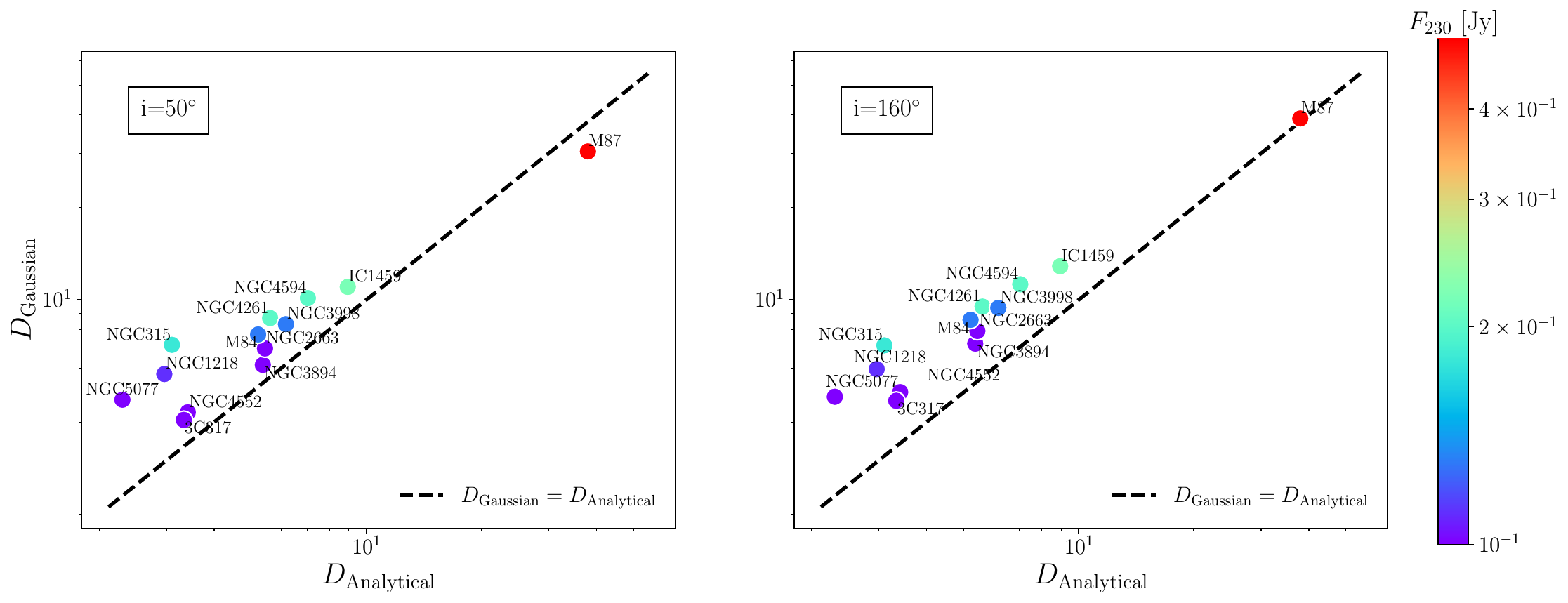}
\caption{Gaussian fit sources sizes as a function of photon ring diameter for $i=50^\circ$ (left) and $i=160^\circ$ (right), based solely on short baseline visibility amplitudes.  We find that sizes derived from Gaussian fitting are 40\% larger on average, since the emitting material has slightly larger angular extent than the photon ring.}
\label{fig:gaussian_fit}
\end{figure*}

However, with precise enough measurements, it is possible to constrain the sizes of our sources with a solely ground-based array by extrapolating the falloff of the visibility amplitudes on short baselines as a function of $(u,v)$ distance \citep{Issaoun+2019}.  We find that simple Gaussian size constraints recover accurate source sizes to within a factor of two, which means that the sources are optically thin enough to produce images comparable in size to the photon ring at these frequencies. To demonstrate this, we perform Gaussian fits in visibility space\footnote{Using the function ``fit\_gauss\_empirical'' in \texttt{eht-imaging} \citep{Chael+2022}.} on the time-averaged total intensity images of our sources at 230 GHz.  Note that noise and realistic uv-sampling are neglected and will be explored in future work.  The results are shown in \autoref{fig:gaussian_fit}, where $D_\mathrm{Gaussian}$ is the mean of the short and long axis FWHM values, and $D_\mathrm{Analytical}$ is the photon ring diameter from \autoref{eqn:photon_ring}.  $D_\mathrm{Gaussian}$ is systematically greater than $D_\mathrm{Analytical}$ by a factor of 40\% on average.  Overestimation is expected, since the emission is clearly of larger angular extent than the photon ring (\autoref{fig:gallery} and \autoref{fig:gallery2}), motivating GRMHD-based calibration as in previous works to more accurately estimate SMBH masses from ring sizes \citep{EHTC+2019f,EHTC+2022f}.  This level of precision is competitive with even direct dynamical mass measurements: recall that the different dynamical mass measurements of our sources in the literature vary by 0.3 dex on average.  Geometric ring model fitting is likely to perform even better \citep{Pesce+2022}.  Note, however, that extended jet emission that is not included in these models may affect these measurements, which should be investigated in future work.  This may manifest as an offset or more rapidly declining component at the shortest baselines in (u,v)-space as observed for \m87 \citep[see Figure 2 of][]{EHTC+2019a}.

\bibliography{main}{}
\bibliographystyle{aasjournal}

\end{document}